\documentclass[preprint]{aastex631}
\usepackage[figuresright]{rotating}
\usepackage{booktabs,makecell,multirow}
\usepackage{bm}
\usepackage{amsmath,amsfonts,amssymb}
\usepackage{array,booktabs,ragged2e}
\usepackage{CJK}

\begin{document}
\begin{CJK*}{UTF8}{gbsn}
                                  
\title{Evolved Massive Stars at Low-metallicity \uppercase\expandafter{\romannumeral6}. Mass-Loss Rate of Red Supergiant Stars in the Large Magellanic Cloud}

\author[0009-0002-6102-579X]{Jing Wen (文静)}
\affiliation{Institute for Frontiers in Astronomy and Astrophysics, Beijing Normal University, Beijing 102206, People's Republic of China}
\affiliation {South-Western Institute for Astronomy Research, Yunnan University, Kunming 650500, People's Republic of China}
\affiliation{Department of Astronomy, Beijing Normal University, Beijing 100875, People's Republic of China}

\correspondingauthor{Jian Gao, Ming Yang, Bingqiu Chen}
\email{jiangao@bnu.edu.cn, myang@nao.cas.cn, bchen@ynu.edu.cn}

\author[0000-0003-4195-0195]{Jian Gao (高健)}
\affiliation{Institute for Frontiers in Astronomy and Astrophysics, Beijing Normal University, Beijing 102206, People's Republic of China}
\affiliation{Department of Astronomy, Beijing Normal University, Beijing 100875, People's Republic of China}

\author[0000-0001-8247-4936]{ Ming Yang (杨明) }
\affiliation{Key Laboratory of Space Astronomy and Technology, National Astronomical Observatories, Chinese Academy of Sciences, Beijing 100101, People's Republic of China}

\author[0000-0003-2472-4903]{Bingqiu Chen (陈丙秋)}
\affiliation{South-Western Institute for Astronomy Research, Yunnan University, Kunming 650500, People's Republic of China}

\author[0000-0003-1218-8699]{Yi Ren (任逸)}
\affiliation{College of Physics and Electronic Engineering, Qilu Normal University, Jinan 250200, People's Republic of China }

\author[0000-0001-5197-4858]{Tianding Wang (王天丁)}
\affiliation{Department of Astronomy, Beijing Normal University, Beijing 100875, People's Republic of China}
\author[0000-0003-3168-2617]{Biwei Jiang (姜碧沩)}
\affiliation{Institute for Frontiers in Astronomy and Astrophysics, Beijing Normal University, Beijing 102206, People's Republic of China}
\affiliation{Department of Astronomy, Beijing Normal University, Beijing 100875, People's Republic of China}



\begin{abstract}

Mass loss is a crucial process that affects the observational properties, evolution path and fate of highly evolved stars. However, the mechanism of mass loss is still unclear, and the mass-loss rate (MLR) of red supergiant stars (RSGs) requires further research and precise evaluation. To address this, we utilized an updated and complete sample of RSGs in the Large Magellanic Cloud (LMC) and employed the 2-DUST radiation transfer model and spectral energy distribution fitting approach to determine the dust-production rates (DPRs) and dust properties of the RSGs. We have fitted 4,714 selected RSGs with over 100,000 theoretical templates of evolved stars. Our results show that the DPR range  of RSGs in the LMC is $10^{-11}\, \rm{M_{\sun}\, yr^{-1}}$ to $10^{-7}\, \rm{M_{\sun}\, yr^{-1}}$, and the total DPR of all RSGs is 1.14 $\times 10^{-6} \, \rm{M_{\sun} \, yr^{-1}}$. We find that 63.3\% RSGs are oxygen-rich, and they account for 97.2\% of the total DPR. The optically thin RSG, which comprise 30.6\% of our sample, contribute only 0.1\% of the total DPR, while carbon-rich RSGs (6.1\%) produce 2.7\% of the total DPR. Overall, 208 RSGs contributed 76.6\% of the total DPR. We have established a new relationship between the MLR and luminosity of RSGs in the LMC, which exhibits a positive trend and a clear turning point at $\log{L/L_{\sun}} \approx 4.4$.
\end{abstract}
\keywords{Red supergiant stars (1375); Stellar mass loss (1613); Circumstellar dust (236); Interstellar medium(847); Large Magellanic Cloud(903)}


\section{Introduction} \label{sec:intro}
Stars with initial masses between 7 and 30\,$\rm M_{\sun}$ follow a path of evolution that leads them to become red supergiants (RSGs) before ultimately ending their lives as core collapse supernovae \citep{2009MNRAS.395.1409S}. During the RSG phase, these stars have low effective temperatures ranging from 3,500 to 4,500\,K, high luminosities from 4,000 to 400,000 $\rm{L_{\sun}}$, and large radii from 100 to 1,500 $\rm{R_{\sun}}$ \citep{1998ApJ...505..793M,2003AJ....126.2867M,2010NewAR..54....1L,2017ars..book.....L}. For stars with initial masses below 30 \,$\rm{M_{\sun}}$, their mass loss mainly occurs during the RSG phase \citep{2020MNRAS.492.5994B}. The type of supernovae that a massive star will become is primarily determined by its initial mass, metallicity, as well as its mass-loss rate (MLR) during the RSG phase  \citep{2010ASPC..425..247H,2013EAS....60...31E,2014ARA&A..52..487S,2015A&A...575A..60M,2018MNRAS.475...55B}. Moreover, RSGs are important contributors to the interstellar medium. The mass loss and evolution of RSGs will eventually influence the chemical evolution of their surrounding mediums. However, the mass loss mechanism of RSGs, which is the result of a combination of factors including radiation pressure, stellar pulsation, stellar wind, convection, as well as the luminosity and metallicity of the stars , is not yet fully understood \citep{1971ApJ...165..285G,1992ApJ...397..644M,2001ApJ...551.1073H,2000ApJ...542..120W,2010ApJ...717L..62Y,2011A&A...526A.156M,2016MNRAS.463.1269B}.

Precisely determining the MLRs or dust-production rates (DPRs) of RSGs is necessary to investigate their mass-loss mechanism. One commonly used method to obtain the MLRs of RSGs is to fit the observational optical to mid-infrared (MIR) spectral energy distributions (SEDs) of the RSGs with theoretical models. Using the DUSTY radiative-transfer code \citep{1997MNRAS.287..799I}, \citet[hereafter W+21]{2021ApJ...912..112W} found an average MLR of $\sim$ 2.0 $\times 10^{-5}\, \rm{M_{\sun} \, yr^{-1}}$ for 1,741 and 1,983 RSGs in M31 and M33. \citet[hereafter Y+23]{2023A&A...676A..84Y}found that the typical MLR of RSGs in the Small Magellanic Cloud (SMC) is about $10^{-6}\, \rm{M_{\sun}\, yr^{-1}}$. \citet{2018AJ....155..212G} estimated the MLR of some well-known RSGs to be ranging from $\times 10^{-4} \, \rm{M_{\sun} \, yr^{-1}}$ to $\times 10^{-6} \, \rm{M_{\sun} \, yr^{-1}}$. \citet[hereafter R+12]{2012ApJ...753...71R} obtained DPRs based on 30,000 RSGs and asymptotic giant branch stars (AGBs) in the Large Magellanic Cloud (LMC) using the GRAMS model grid \citep{2011A&A...532A..54S,2011ApJ...728...93S}, which were mostly between $10^{-7}$ and $10^{-11} \, \rm{M_{\sun} \, yr^{-1}}$.  \citet[hereafter B+12]{2012ApJ...748...40B} calculated the DPRs of AGBs and RSGs in the LMC and SMC from their infrared excesses and obtained a total DPR of approximately $2.4\times 10^{-7} \, \rm{M_{\sun}\, yr^{-1}}$ for 3,908 RSGs in the LMC. \citet[hereafter S+16]{2016MNRAS.457.2814S}  conducted a study on the DPR of 1,410 RSGs in the SMC, which had a total DPR of $4.6\times 10^{-8} \, \rm{M_{\sun}\, yr^{-1}}$ and an average DPR of $3.3\times 10^{-11} \, \rm{M_{\sun}\, yr^{-1}}$. The MLRs obtained from these works vary significantly, mainly due to the use of different dust models and possible bias of the adopted samples sizes \citep{2016MNRAS.457.2814S,2018A&A...609A.114G}. A complete and pure RSG sample is vital for us to accurately calculate the MLRs of RSGs.

\citet{2021ApJ...923..232R} constructed comprehensive and uncontaminated RSG samples of Local Group galaxies including the LMC and SMC, which offers the oppotunity to investigate RSGs mass loss in the LMC and their overall contribution to interstellar medium. The following sections of this paper are organized as follows: Section~\ref{sec:samples} details the sample, Section~\ref{sec:models} presents the dust model grid, Section~\ref{sec:fitting} explains the fitting procedure, Section~\ref{sec:result} shows the results of this research, and Section~\ref{sec:summary} summarizes the paper.

\section{Data}
\label{sec:samples}
In the current work, we adopt the RSG sample from \citet{2021ApJ...923..232R}, who utilized the near-infrared photometric data from
UKIRT/WFCAM \citep{2013ASSP...37..229I} and the 2MASS point source catalog \citep{2006AJ....131.1163S} and astrometric data from Gaia EDR3  \citep{2021A&A...649A...1G} to identify RSGs in 12 low-mass galaxies of the Local Group. In their sample, there are 4,823 RSGs in the LMC.

In order to estimate MLRs of the RSGs, SEDs with wide coverage of wavelength range is necessary. The infrared (IR) photometry is essential as it reveals information about the circumstellar dust, while the optical data provide atmospheric parameters of the star.  We crossmatched the RSG sample with various photometric data, including Skymapper (\textit{u}, \textit{v}, \textit{g}, \textit{i}, \textit{r}, and \textit{z} bands; \citealt{2007PASA...24....1K,2011PASP..123..789B,2018PASA...35...10W}), Gaia EDR3 ($G_{\rm{BP}}$, \textit{G}, and $G_{\rm{RP}}$ bands; \citealt{2021A&A...649A...1G}), Magellanic Clouds Photometric Survey (MCPS; \textit{U}, \textit{B}, \textit{V}, and \textit{I} bands; \citealt{2004AJ....128.1606Z}),  Vista Magellanic Cloud Survey (VMC; \textit{Y}, \textit{J}, and $K_{\rm{S}}$ bands; \citealt{2011A&A...527A.116C}), InfraRed Survey Facility (IRSF; \textit{J}, \textit{H}, and $K_{\rm{S}}$ bands; \citealt{2007PASJ...59..615K}), The Two Micron All Sky Survey (2MASS; \textit{J}, \textit{H}, and $K_{\rm{S}}$ band; \citealt{2006AJ....131.1163S}), AKARI (\textit{N3}, \textit{S7}, \textit{S11}, \textit{L15}, and \textit{L24} bands; \citealt{2007PASJ...59S.401O,2007PASJ...59S.369M,2012AJ....144..179K}), Wide field Infrared Survey Explorer (WISE; \textit{W1}, \textit{W2}, \textit{W3}, and \textit{W4} bands; \citealt{2010AJ....140.1868W}), and Spitzer ([3.6], [4.5], [5.8], [8.0], and [24] bands; \citealt{2004ApJS..154....1W}), as shown in Table~\ref{tab:data}. A search radius of $1^{\arcsec}$ was used in the crossmatching. These photometric data provided a broad wavelength coverage ranging from ultraviolet (0.36\,$\mu m$) to IR (24\,$\mu m$). To ensure data quality, we also imposed additional constraints on the photometric data for each band as following:

\begin{enumerate}
\item{For Skymapper, Gaia, VMC, 2MASS, IRSF, and AKARI data, we required that the magnitude errors must be less than 0.1 mag.}
\item{For WISE data, we required that the $W1$ and $W2$ magnitude errors must be less than 0.2 mag, and $W3$ and $W4$ magnitude errors must be less than 0.3 mag. We also constrained the signal-to-noise ratio (SNR) as $\rm{SNR}_\textit{W1} \geq 5, \rm{SNR}_\textit{W2} \geq 5, \rm{SNR}_\textit{W3} \geq 7, \rm{SNR}_\textit{W4} \geq 10$, and $ex\_flag \footnote{The probability that the source form is inconsistent with a single PSF, as detailed in \href{https://irsa.ipac.caltech.edu/cgi-bin/Gator/nph-scan?mission=irsa\&submit=Select\&projshort=WISE}{https://irsa.ipac.caltech.edu/cgi-bin/Gator/nph-scan?mission=irsa\&submit=Select\&projshort=WISE}.}= 0$, $ nb\footnote{The number of blend components used in each fit.} = 1$, $w1/2/3/4cc\_map\footnote{Contamination and confusion bit array for the source in corresponding band.} = 0$ or $w1/2/3/4flg \footnote{Instrumental standard aperture flag.}=0$}.
\item {For Spitzer data, we require the SNR of each band to be no less than 3 and the close source flag $close\_flag = 0$.}
\end{enumerate}

 Meanwhile, it was not necessarily for each target in our sample to have observations in all filters and satisfy above conditions, which would significantly reduced the number of stars in our sample. However, to ensure relatively accurate estimation of MLRs, 2MASS $JHK_{\rm{S}}$ measurements for each star must be good. We also required that each source must had good observations in WISE $W3$ or Spitzer [8.0] band, which was important to constrain the dust emission information. After applying these criteria, 85 sources were excluded based on photometric quality constraints. Among these 85 sources, one did not meet the 2MASS criteria , while the remaining 84 lacked data meeting the selection criteria for W3 or [8.0]. We performed cross-matching with Simbad \citep{2000A&AS..143....9W} and removed 24 sources that were identified as planetary nebula, post-AGB, or YSO etc. After applying these criteria, there were 4,714 RSGs in the final RSG sample.

{\catcode`\&=11
\gdef\AandA116{\cite{2011A&A...527A.116C}}}
{\catcode`\&=11
\gdef\2021AandA...649A...1G{\cite{2021A&A...649A...1G}}}
\begin{deluxetable*}{cccc}
\tabletypesize{\tiny}
\tablecolumns{4}
\renewcommand\arraystretch{1.3}
\tablecaption{Datasets for the spectral energy distribution fitting\label{tab:data}}
\tablehead{\colhead{Datasets}&\colhead{Filters}&\colhead{Selection criteria}&\colhead{N\%$^{\rm a}$}}
\startdata
 Skymapper$^{\rm b}$ & $u$, $v$, $g$, $i$, $r$, $z$ & $e\_u/v/g/i/r/zPSF<0.1$   & 30\%, 27\%, 97\%, 97\%, 97\%, 98\% \\
MCPS $^{\rm c}$ & \textit{U}, \textit{B}, \textit{V}, \textit{I}, & $e\_U/B/V/Imag<0.1$   & 76\%, 84\%, 84\%, 57\%\\
 Gaia EDR3 $^{\rm d}$&$G_{\rm{BP}}$, \textit{G}, $G_{\rm{RP}}$&$phot\_bp/g/rp\_mean\_mag\_error < 0.1$&98\%, 100\%, 99\%\\
VMC$^{\rm e}$&\textit{Y}, \textit{J}, $K_{\rm{S}}$&$e\_Y/J/K_{\rm S}pmag < 0.1$& 26\%, 26\%, 26\%\\
IRSF$^{\rm f}$&\textit{J}, \textit{H}, $K_{\rm{S}}$&$e\_J/H/K_{\rm{S}}mag\_IRSF<0.1$&83\%, 75\%, 62\%\\ 
2MASS$^{\rm g}$&\textit{J}, \textit{H}, $K_{\rm{S}}$&$e\_J/H/K_{\rm S}mag < 0.1$& 100\%, 100\%, 100\%\\
AKARI$^{\rm h}$& \textit{N3}, \textit{S7}, \textit{S11}, \textit{L15}, \textit{L24}&$e\_N3/S7/S11/L15/L24mag<0.1$&28\%, 25\%,  23\%, 7\%, 4\%\\
\multirow{5} *{ WISE$^{\rm i}$}&\multirow{5} *{\textit{W1}, \textit{W2}, \textit{W3}, \textit{W4}}& $e\_W1mag<0.2, e\_W2mag<0.2$& \multirow{5}*{61\%, 60\%, 53\%, 6\%}\\
{ }& { }&$e\_W3mag<0.3, e\_W4mag<0.3$& \\
{ }&{ }&$\rm{SNR}_\textit{W1}\geq 5, \rm{SNR}_\textit{W2}\geq 5$&\\
{ }&{ }&$\rm{SNR}_\textit{W3}\geq 7, \rm{SNR}_\textit{W4}\geq 10$&\\
{ }&{ }&$ex\_flag = 0, nb = 1, w1/2/3/4cc\_map = 0 $ or $w1/2/3/4flg =0$&\\
Spitzer$^{\rm j}$&[3.6], [4.5], [5.8], [8.0], [24]& $\rm{SNR}[3.6]/[4.5]/[5.8]/[8.0]\geq3,closeflag=0$& 94\%, 94\%, 94\%, 94\%, 34\%
\enddata 
\tablecomments{$^{\rm a}$ The proportion of RSGs in this band with respect to the total sample. $^{\rm b}$ 
 \citet{2007PASA...24....1K,2011PASP..123..789B,2018PASA...35...10W} $^{\rm c}$ \citet{2004AJ....128.1606Z} $^{\rm d}$ \2021AandA...649A...1G $^{\rm e}$ 
  \AandA116 $^{\rm f}$ \citet{2007PASJ...59..615K} $^{\rm g}$ \citet{2006AJ....131.1163S} $^{\rm h}$ \citet{2007PASJ...59S.401O,2007PASJ...59S.369M,2012AJ....144..179K} $^{\rm i}$  \citet{2010AJ....140.1868W} $^{\rm j}$ \citet{2004ApJS..154....1W}}
\end{deluxetable*}

\begin{figure}
   \centering
	\includegraphics[width=0.6\linewidth]{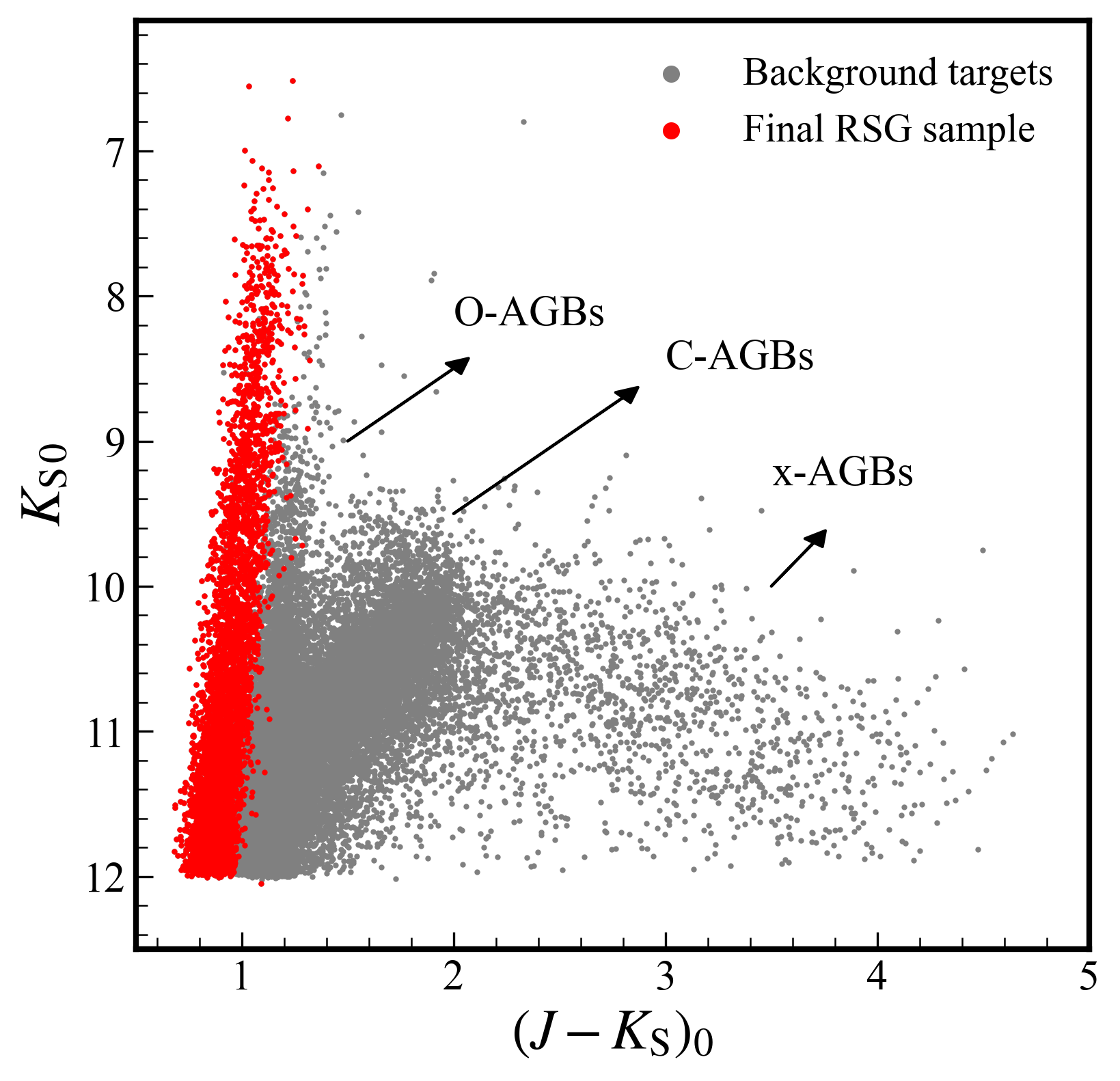}
    \caption{  The 2MASS Color-magnitude diagram including background about 30,000 AGB stars and final 4,714 RSGs (red).}
   \label{fig:cmd}
\end{figure}

We then corrected for the extinction effect of our sample. In the current work, we utilized the extinction map of LMC by \citet{2022MNRAS.511.1317C} and the extinction coefficients from the classical CCM89 model \citep{1989ApJ...345..245C} with $R_{\rm V}= 3.41$ \citep{2003ApJ...594..279G}.
In our sample, there are 8 stars are not covered by the extinction map of \citet{2022MNRAS.511.1317C} and another 155 stars have reddening values $E(B-V) < 0 $ in the map, which could be due to minimal extinction and errors. We used the same method as Y+23 to correct the extinction of these 163 sources, setting $R_{\rm V}=3.1$, $E(B-V)=0.1$ \citep{2011ApJ...737..103S,2020ApJ...891...57F}, adopting the extinction coefficient obtained by \citet{2019ApJ...877..116W}.The 2MASS $J$ and $K_{\rm S}$ bands color-magnitude diagram (CMD) of the background targets and final RSG sample after extinction correction are presented in Figure~\ref{fig:cmd}.

\section{Model}
\label{sec:models}
Various radiation transfer models are available for calculating the
MLR, including one-dimensional model DUSTY \citep{1997MNRAS.287..799I}, two-dimensional model 2-DUST \citep{2003ApJ...586.1338U}, and three-dimensional model MCMax3D \citep{2009A&A...497..155M}, etc. Y+23 used the 1D spherically symmetric DUSTY radiative transfer model and identified a prominent turning point on the luminosity-MLR diagram for RSGs. We speculate that this turning point may be related to the mass loss mechanism of RSGs  \citep[e.g.,][]{2023A&A...678L...3V}. To confirm the presence of this turning point and determine if it is influenced by factors such as models and samples, we decided to employ different radiative transfer models. We adopted 2-DUST to generate the theoretical SEDs. 2-DUST is a flexible, axisymmetric 2D radiation transfer code with highly adaptable input parameters. We assumed a spherically symmetric circumstellar dust shell for our study. With input parameters such as the spectrum of the central star, dust shell geometry, and dust grain properties, 2-DUST solves the radiation transfer problem and provides DPR, luminosity, and other output parameters of the star \citep{2003ApJ...586.1338U}. The settings of our input parameters are outlined in Table ~\ref{tab:model}. We also provided the parameter settings of GRAMS \citep{2011ApJ...728...93S,2011A&A...532A..54S}, a template library generated using the 2-DUST radiation transfer model, for comparison. The input parameters we used in 2-DUST can be divided into three parts: the spectrum of the central star, the shape of the dust shell, and the dust grain properties of the circumstellar dust. We described them in detail below. 

{\catcode`\&=11
\gdef\AandAS{\cite{2011A&A...532A..54S}}}
{\catcode`\&=11
\gdef\AandAG{\cite{2008A&A...486..951G}}}
{\catcode`\&=11
\gdef\AandAK{\cite{2005A&A...442..281K}}}
{\catcode`\&=11
\gdef\AandAKK{\cite{2006A&A...452.1021K}}}
{\catcode`\&=11
\gdef\AandAD{\cite{1995A&A...300..503D}}}
{\catcode`\&=11
\gdef\AandAP{\cite{1988A&A...194..335P}}}
{\catcode`\&=11
\gdef\AandAAA{\cite{2009A&A...503..913A}}}

\begin{deluxetable}{ccccc}
\caption{\label{tab:model}Parameters for models}
\tabletypesize{\scriptsize}
\tablehead{ Parameter &  C-rich (This work) &  O-rich (This work) & GRAMS C-rich$^{\rm a}$ &  GRAMS O-rich$^{\rm b}$}
\startdata
\textbf{Photosphere model }& MARCS$^{\rm c}$ & MARCS  &COMARCS$^{\rm d}$&PHOENIX$^{\rm e}$\\
$L_{\rm{mod}}\,(\rm{L_{\sun}})$&$1140-485\,000$&$1140-485\,000$&$1100-26\,000$&$1000-1\,000\,000$\\
\textit{$T_{\rm{eff}}\,(\rm{K})$}&2500 $-$4500 & 2500 $-$ 5000 &2600 $-$ 4000& 2100 $-$ 4700 \\
$\log{g}$ &$-0.5 - 0$ & $ -0.5 - 0$ &$-1 - 0$& $-0.5$\\
\textit{M\,($\rm{M_{\sun}}$)}&$0.5 - 15$ &$0.5 - 15$&$ 1 - 5 $& 1\\
\midrule[0.5pt]
\textbf{Dust shell properties}&&\\
density distribution& $\rho(r)\propto r^{-2}$ & $\rho(r) \propto r^{-2}$ & $\rho(r)\propto r^{-2}$& $\rho(r)\propto r^{-2}$\\
\textit{$R_{\rm{in}}\,(R_{\star})$}& 3,5,7,10,12,15,20,25& 3,5,7,10,12,15,20,25 &1.5,3,4.5,7,12&3,7,11,15 \\
\textit{$R_{\rm{out}}\,(R_{\rm{in})}$}& 1000&1000&1000&1000\\
$v_{\rm{exp}}\,(\rm{cm\, s^{-1}})$&10&10&10&10\\
\midrule[0.5pt]
\textbf{Dust grain properties}&&\\
Species&AmC(Zubko+96$^{\rm f}$) & Silicate(DL84$^{\rm g}$,Do+95$^{\rm h}$)&AmC(Zubko+96)+SiC(Pegourie+88$^{\rm i}$) & Silicate(Sargent+10$^{\rm j}$) \\
$\tau_{11.3}$ or $\tau_{9.7}$ & $10^{-4} -  10 $&$ 10^{-4} - 30 $& $10^{-4} - 4$& $10^{-4} - 26$\\
Size distribution& KMH$^{\rm k}$&KMH& KMH&KMH\\
$a_{\rm{min}}\,(\mu m)$&0.01 & 0.01&0.01&0.01\\
$a_{0}\,(\mu m)$& 1& 0.25& 0.1&1\\
$\rho_{\rm d}(\rm{g\, cm^{-3}})$ &1.8 &3.5& 1.8,3.22&3.2\\
$\gamma$&3.5&3.5&3.5&3.5\\
\midrule[0.5pt]
$\dot{M}_{\rm d}$\,($\rm{M_{\sun}\,yr^{-1}}$) &2.4$\times10^{-13}- 1.9\times10^{-6}$&1.7$\times10^{-13}- 1.0\times10^{-6}$ &1.5$\times10^{-12}- 2.1\times10^{-7}$&3$\times10^{-13}- 3\times10^{-5}$\\
\enddata
\tablecomments{$^{\rm a}$ \AandAS $^{\rm b}$ \citet{2011ApJ...728...93S} $^{\rm c}$ \AandAG  $^{\rm d}$\AandAAA  $^{\rm e}$ \AandAK , \AandAKK $^{\rm f}$ \citet{1996MNRAS.282.1321Z} $^{\rm g}$ \citet{1984ApJ...285...89D} $^{\rm h}$ \AandAD  $^{\rm i}$ \AandAP $^{\rm j}$ \citet{2010ApJ...716..878S}} 
    
\end{deluxetable}

\subsection{Photosphere Model of the centeral star}
\begin{figure}
\centering
	\includegraphics[width=0.6\columnwidth]{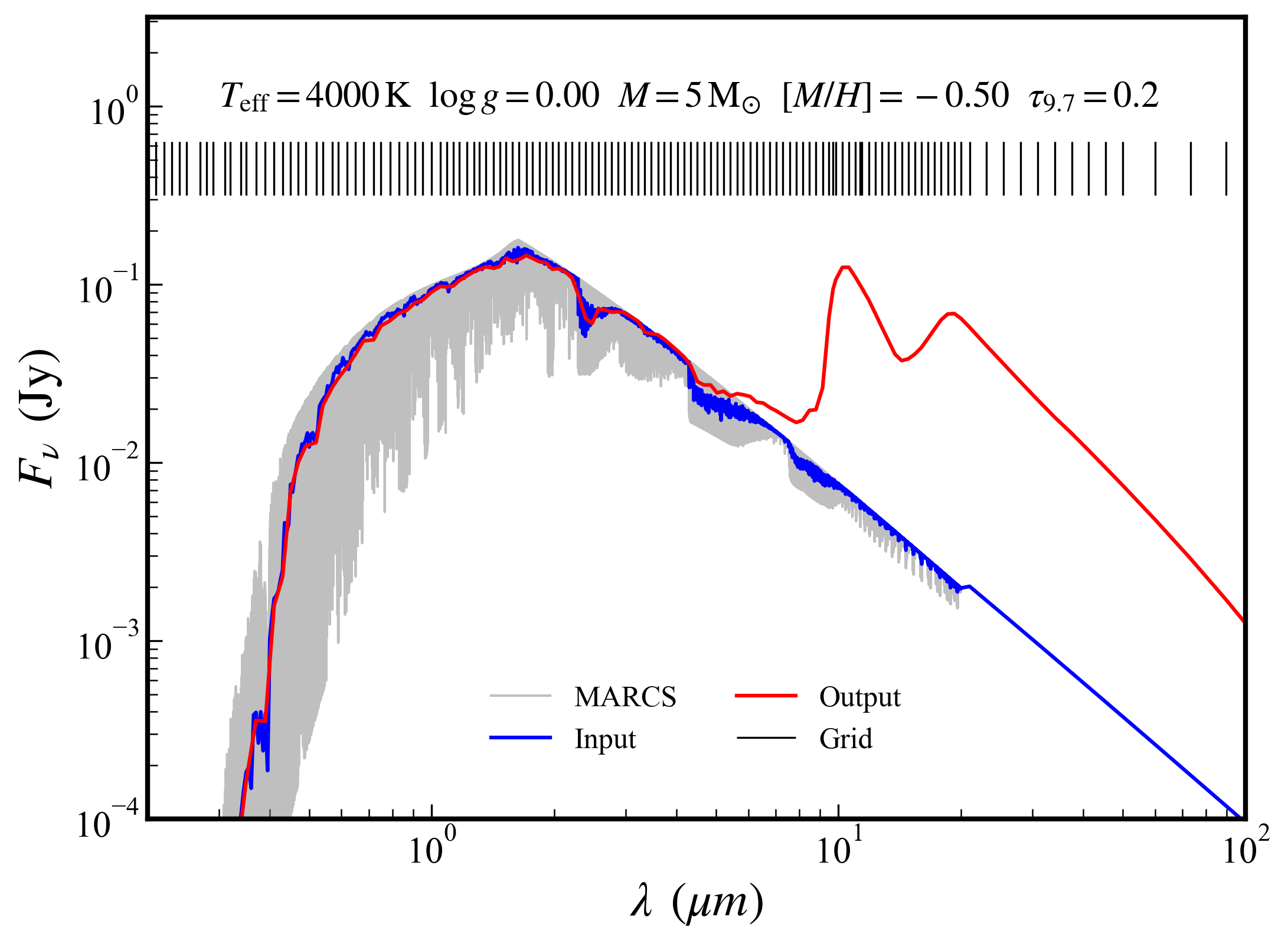}
    \caption{Spectra of an example model with $T_{\rm{eff}}=4,000\,\rm{K}$, $\log{g}=0.00$ and  ${M}=5\,\rm{M_{\odot}}$, ${[M/H]}=-0.50$, $\tau_{9.7}=0.2$. The gray spectrum is the MARCS stellar spectrum. The blue and red spectrum are the input and output spectra of 2-DUST, respectively. The black vertical lines above the spectra represent the output wavelength grid.}
    \label{fig:sed}
\end{figure}

For the spectrum of the central star, we adopted the MARCS
stellar atmosphere model \citep{2008A&A...486..951G}. A blackbody spectrum may be easier to use, but it fails to reflect evolutionary effects
such as the ``H-bump" feature \citep{2021A&A...647A.167Y}. Our aim was to
generate an SED grid that covers the properties of the evolved stars, including both AGB stars and RSGs in the Magellanic Clouds. 40 MARCS models are chosen as templates for RSGs, with temperatures ranging from 3,300 to 4,500\,K (3300, 3400, 3500, 3600, 3700, 3800, 3900, 4000, 4250, 4500\,K), a mass of 15\,$\rm{M_{\sun}}$, surface gravity values ranging from $-0.5$ to 0 dex, and metallicities ranging from $-0.5$ to 0 dex. For evolved stars in other mass ranges, we also used 78 MARCS models with effective temperatures ranging from 2,500 to 5,000\,K (2500, 2600, 2700, 2800, 2900, 3200, 3300, 3400, 3500, 3600, 3700, 3800, 3900, 4000, 4250, 4500, 4750, 5000\,K), masses ranging from 0.5 to 5\,$\rm{M_{\sun}}$ (0.5, 1, 2, 5\,$\rm{M_{\sun}}$), surface gravity values ranging from $-0.5$ to 0\,dex, and a metallicity is $0.5$\,dex. Among them, the MARCS models with a temperature higher than 4500\,K are only used to generate oxygen rich (O-rich) templates.  We extended the wavelength coverage of the MARCS spectra to 1,000 $\mu m$ by adding blackbody in the infrared region ( $\geq 20 \mu m$). The MARCS models were then downgraded to a resolution of $R=100$ and fed into 2-DUST. To reduce the computation time, we set the number of data points output by the 2-DUST to 142. Figure~\ref{fig:sed} shows the 2-DUST input and output spectra of an example model.

\subsection{Circumstellar Dust Shell}
We assumed that the density distribution of the dust envelope followed a power-law relation of $\rho(r)\propto r^{-2}$, and that the
dust envelope was spherically symmetric. For templates with masses
between 0.5 and 5\,$\rm{M_{\sun}}$, we calculated the model with inner dust shell radii of $R_{\rm{in}}=$ 3, 5, 7, 10, 12, 15$\,R_{\star}$, and for those with mass of 15\,$\rm{M_{\sun}}$, $R_{\rm{in}}=$5, 7, 10, 15, 20, 25$\,R_{\star}$, for which $R_{\star}$ was the radius of the central star. The ratio between the outer radius and inner radius was assumed to be $R_{\rm{out}}/R_{\rm{in}}= 1000$.

The temperature of the inner dust shell was determined by its size in 2-DUST. Previous works estimated the condensation temperature of silicate dust to be around 400 to 1,500\,K, which was influenced by the central star and the dust composition \citep{1984A&A...132..163G,1999A&A...347..594G,2005A&A...438..273V,2010ApJ...716..878S,2020A&A...644A.139G}. On the other hand, the condensation temperature of carbon dust is higher \citep{1994A&A...287..163G,2013MNRAS.434.2390N,2018A&A...614A..17B}, and the upper limit of carbon dust temperature in GRAMS is set to 1800 K \citep{2011A&A...532A..54S}. For our case, we used 1,800 K as the upper limit for carbon-rich (C-rich) models and 1,400 K for O-rich models.

The calculation of DPR by 2-DUST requires wind speed, which is difficult to measure. Previous studies of evolved AGB and RSG stars obtained a wide velocity range \citep{1999A&A...351..559V,2006A&A...454L.103R,2010A&A...523A..18D,2012A&A...537A..35C,2004MNRAS.355.1348M}. We assumed a constant expansion velocity ($v_{\rm{exp}}$) of $v_{\rm{exp}}= 10 \rm{\,km\,s^{-1}}$ without considering the influence of luminosity and metallicity, etc. $v_{\rm{exp}}$ and DPR are simply proportional, that different DPR can be obtained by changing the values of wind speeds after all grids are generated. We will discuss later (Section~\ref{subsec:result2}) the result of scaling $v_{\rm{exp}}$ and $v_{\rm{exp}}= 25 \rm{\,km\,s^{-1}}$.

\subsection{Circumstellar Dust Grains}
The chemical composition of RSG wind was mainly O-rich, identified by the wide, smooth, and featureless Si-O stretching 9.7 $\mu m$ and O-Si-O bending 18 $\mu m$ silicate features. For dust grains, we assumed that the composition of the O-rich model circumstellar dust was amorphous silicate with optical constants adopted from \citet{1984ApJ...285...89D} and a density of 3.5\,$\rm{g \,cm^{-3}}$. Some RSGs were also found to have the carbonaceous features \citep{1998MNRAS.301.1083S,2009A&A...498..127V}. Thus, we assumed the composition of C-rich model dust was the common amorphous carbon (AMC) without silicon carbide (SiC) and polycyclic aromatic hydrocarbons (PAHs). The density of AMC was 1.8\,$\rm{g \,cm^{-3}}$ with optical constant came from \citet{1996MNRAS.282.1321Z}. We used the KMH size distribution of dust grains from \citet{1994ApJ...422..164K}, where the power-law index of $\gamma = 3.5$, the minimum grain size of $a_{\rm{min}} = 0.01\,\mu m$, and the exponential scale height of $a_{0} = 1\,\mu m$ for C-rich dust and $a_{0} = 0.25\,\mu m$ for O-rich dust, respectively. For the optical depth $\tau_{\lambda}$, we assumed $\lambda = 9.7 \,\mu m$ for the O-rich model and 11.3 $\,\mu m$ for the C-rich model, respectively \citep{2011ApJ...728...93S,2011A&A...532A..54S}. We set 64 steps of optical depth for $\tau_{11.3}$, ranging from 0.0001 to 10 (C-rich models), and 84 steps of $\tau_{9.7}$, ranging form 0.0001 to 30 (O-rich models), with steps of 0.00018 for $\tau_{\rm 9.7/11.3}= 0.0001 \sim 0.001$, 0.001 for $\tau_{\rm 9.7/11.3}= 0.001 \sim 0.01$, 0.005 for $\tau_{\rm 9.7/11.3}= 0.01 \sim 0.1$, 0.05 for $\tau_{\rm 9.7/11.3}= 0.1 \sim 1$, 0.5 for $\tau_{\rm 9.7/11.3}=1 \sim 5$ and 1 for $\tau_{\rm 11.3} = 5 \sim 10$, 1 for $\tau_{\rm 9.7}=5 \sim 30$.

Based on above configuration, we generated a total of more than 10,000 models. We selected models with a dust shell inner temperature lower than 1,400\,K (O-rich) and 1,800\,K (C-rich), which yielded 77,458 models, consisting of 35,159 C-rich models and 42,299 O-rich models. To calculate the luminosity of each target, we assumed a distance of 50\,kpc for the LMC \citep{1997macl.book.....W,2011ApJS..192....6L}. The luminosity was estimated in three ways: using the 2-DUST output, the integrated observational SED ($L_{\rm{obs}}$), or the integrated best fitted model SED ($L_{\rm{mod}}$). The outputs for all the model are listed in Table~\ref{tab:model2}, this table is available in the machine-readable format, a portion is shown here for guidance regarding its form and content.

\movetabledown=2in
\begin{rotatetable}
\begin{deluxetable}{cccccccccccccccccccc}
\tabletypesize{\scriptsize}
\tablewidth{0pt}
\tablecaption{Template library of evolved stars\label{tab:model2}}
\tablehead{
\colhead{ID} &  \multicolumn{3}{c}{SED}  & \multicolumn{3}{c}{Simulated photometric data}   & \multicolumn{6}{c}{Dust shell parameters}&   \multicolumn{7}{c}{Center star parameters}\\
\cmidrule(lr){1-1}\cmidrule(lr){2-4}\cmidrule(lr){5-7}\cmidrule(lr){8-13}\cmidrule(lr){14-20} 
index& SED0.2$^{\rm a}$ & ...... & SED1000 & mfluxsku$^{\rm b}$ & ...... & mfluxspitzer24 &$\log{\dot{M}_{\rm d}} $ & $\tau$ & $R_{\rm in}$$^{\rm c}$ & $T_{\rm d}^{\rm d}$ & type & ${\rho_{\rm in}}^{\rm e}$ & tmass$^{\rm f}$ & $\log{L_{\rm mod}}$ & $T_{\rm{eff}}$ &  Mass$^{\rm g}$ & Z & $\log{g}$\\
&(Jy) & ...... & (Jy) & (Jy) & ...... & (Jy) & ($ \rm{M_{\sun} \, yr^{-1}}$) & &($R_{\star}$)& (K)& &($\rm{g\,cm^{-3}}$) &($\rm{M_{\sun}}$) &($\rm L_{\sun}$) &(K) &($\rm{M_{\sun}}$)& &&}
\startdata 
        0 & 8.46E-12 & ...... & 2.44E-06 & 6.20E-05 & ...... & 4.13E-03 & -11.058  & 0.0001 & 20 & 711  & C & 5.48E-23 & 2.48E-07 & 4.647  & 3300 & 15 & 0 & 0 \\ 
        1 & 8.46E-12 & ...... & 2.44E-06 & 6.20E-05 & ...... & 4.17E-03 & -10.963  & 0.0001 & 25 & 652  & C & 4.38E-23 & 3.87E-07 & 4.647  & 3300 & 15 & 0 & 0 \\ 
        2 & 7.11E-10 & ...... & 3.14E-06 & 1.20E-03 & ...... & 5.64E-03 & -10.963  & 0.0001 & 25 & 843  & C & 4.38E-23 & 3.87E-07 & 5.086  & 4250 & 15 & 0 & 0 \\ 
        3 & 3.64E-10 & ...... & 2.89E-06 & 3.22E-04 & ...... & 5.04E-03 & -11.058  & 0.0001 & 20 & 847  & C & 5.48E-23 & 2.48E-07 & 4.937  & 3900 & 15 & 0 & 0 \\ 
        4 & 3.64E-10 & ...... & 2.88E-06 & 3.22E-04 & ...... & 4.97E-03 & -11.183  & 0.0001 & 15 & 949  & C & 7.30E-23 & 1.39E-07 & 4.937  & 3900 & 15 & 0 & 0 \\ 
        5 & 8.46E-12 & ...... & 2.43E-06 & 6.20E-05 & ...... & 4.01E-03 & -11.514  & 0.0001 & 7 & 1081  & C & 1.56E-22 & 3.03E-08 & 4.647  & 3300 & 15 & 0 & 0 \\ 
        6 & 1.67E-10 & ...... & 2.67E-06 & 1.37E-04 & ...... & 4.49E-03 & -11.360  & 0.0001 & 10 & 1028  & C & 1.10E-22 & 6.19E-08 & 4.798  & 3600 & 15 & 0 & 0 \\ 
        7 & 1.67E-10 & ...... & 2.67E-06 & 1.37E-04 & ...... & 4.45E-03 & -11.514  & 0.0001 & 7 & 1187  & C & 1.56E-22 & 3.03E-08 & 4.798  & 3600 & 15 & 0 & 0 \\ 
        8 & 1.67E-10 & ...... & 2.67E-06 & 1.37E-04 & ...... & 4.42E-03 & -11.660  & 0.0001 & 5 & 1360  & C & 2.19E-22 & 1.55E-08 & 4.798  & 3600 & 15 & 0 & 0 \\ 
        9 & 8.46E-12 & ...... & 2.43E-06 & 6.20E-05 & ...... & 4.04E-03 & -11.360  & 0.0001 & 10 & 937  & C & 1.10E-22 & 6.19E-08 & 4.647  & 3300 & 15 & 0 & 0 \\ 
        ...... & ...... & ...... & ...... & ...... & ...... & ...... & ...... & ...... & ...... & ...... & ...... & ...... & ...... & ...... & ...... & ...... & ...... & ...... \\ 
\enddata
\tablecomments{This table is published in its entirety in the machine-readable format. A portion is shown here for guidance regarding its form and content. \\$^{\rm a}$The SED of the model covers a range of 0.2  to 1000 $\,\mu m$, with 142 grid points, which include dust emission and center star. \\ $^{\rm b}$ The synthetic flux calculated by the convolution of the 2-DUST model flux and the individual filters, we adopt the filter parameters from the Spanish Virtual Observatory (SVO) Filter Profile Service, see \href{ http://svo2.cab.inta-csic.es/theory/fps/}{ http://svo2.cab.inta-csic.es/theory/fps/}.\\ $^{\rm c}$ The inner radius of dust shell, measured in units of the radius of the central star. \\$^{\rm d}$Temperature at the inner radius of the dust shell. \\$^{\rm e}$ Density at the inner radius of the dust shell.\\ $^{\rm f}$The mass of the entire dust shell. \\$^{\rm g}$The mass of the central star. }

\end{deluxetable}
\end{rotatetable}

\section{SED Fitting}
\label{sec:fitting}
\begin{figure}
    \centering
	\includegraphics[width=0.5\linewidth]{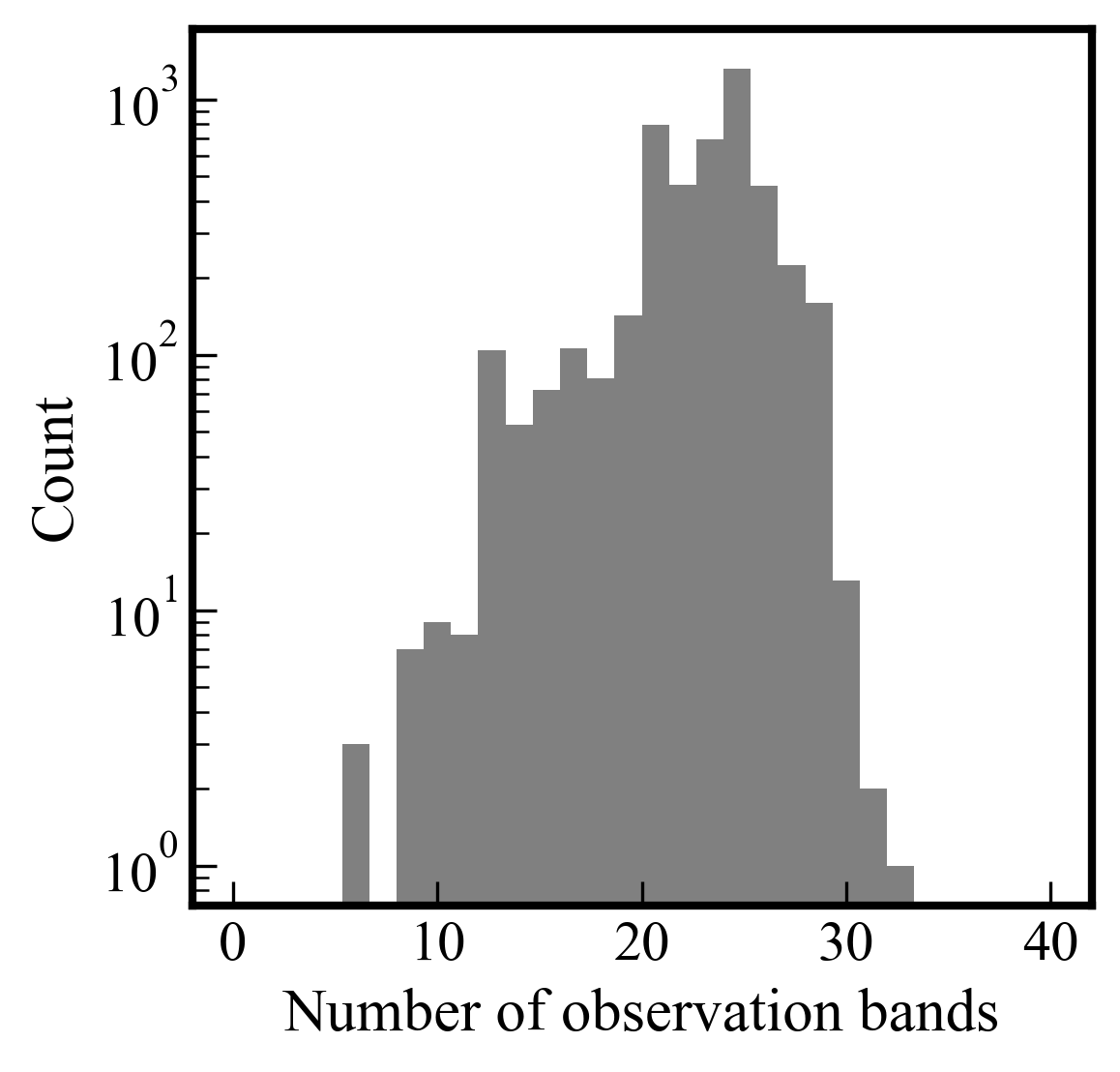}
    \caption{Histogram of numbers of data points for the RSG sample.}
    \label{fig:datacount}
\end{figure}

We determined the best-fit model by minimizing the $\chi^{2}$ value, $\chi^{2}$ for each source was defined by, 

\begin{equation}
\label{eq:chi}
    \chi^2_i=\sum^N\dfrac{C[\log{F(M_i, \lambda)}-\log{F(O, \lambda)}]^2}{N\left|f(O, \lambda)\right|},
\end{equation}

where $i$ is the number of the model, $F(M, \lambda)$ are the simulated photometric data derived by the convolution of the model flux and each individual filters\footnote{We adopted the filter response curves from the Spanish Virtual Observatory (SVO) Filter Profile Service \href{http://svo2.cab.inta-csic.es/theory/fps/}{ (http://svo2.cab.inta-csic.es/theory/fps/)}}, $F(O, \lambda$) are the observed photometric data, $N\left|f(O, \lambda)\right|$ are the number of adopted observation bands, and different for each source. $C$ are the weight used for data of different wavelengths. In order to increase the weight of the infrared data, the C is changing based on different wavelengths, e.g., $C = 5$ for $\lambda \geq 1.0\,\mu m $ and $ C =1 $ for $\lambda < 1.0\,\mu m$. The distribution of the number of measurements, $N\left|f(O, \lambda)\right|$, ranging from 6 to 32 filters for our RSG sample, was shown in Figure~\ref{fig:datacount}. 

\begin{figure}
    \centering
	\includegraphics[width=0.6\linewidth]{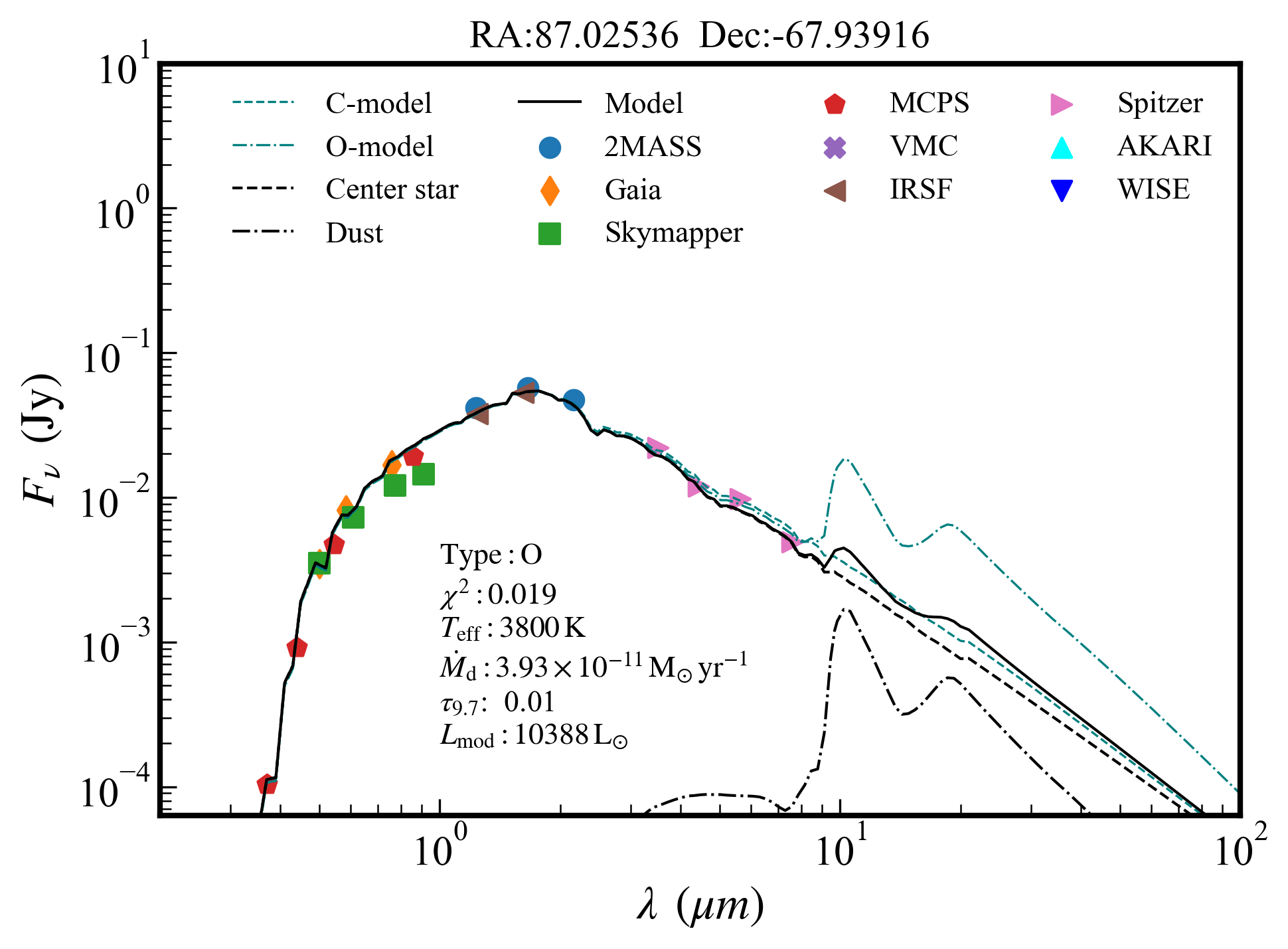}
    \caption{The result of direct fitting without restriction at long wavelengths (blue dot-dashed line) and the result of refitting after the restriction (black solid line) for an example RSG in the S1 sample. The observations are marked with different symbols of different colors. The black solid curves are the best fitted models. The black dashed and dot-dashed lines are the spectra contributed by the center star and the circumstellar dust, respectively. The blue dashed line represents the minimum $\chi^2$ model of another chemical component.}
    \label{fig:fitre}
\end{figure}

Prior to fitting, all the sample RSGs are split into two groups. The first group, named ``S1", is consists of RSGs lacking \textit{W3} data or photometric data with wavelength longer than the \textit{W3} band. The second group, named ``S2", is comprised of RSGs containing measurements of wavelength no less than the \textit{W3} band (i.e. the WISE $W3$ and $W4$, Spitzer [24], AKARI $S11$, $L15$ or AKARI $L24$ band). Stars in the S1 sample can only be fitted with models where $\tau_{9.7/11.3}\leq 0.01$, while those in the S2 sample can be fitted with all models. Photometric data in those long wavelength are essential in determining the dust species present in a circumstellar dust shell (specifically, Si-O stretching $\sim 9.7\,\mu m$ and O-Si-O bending $\sim 18\,\mu m$). The number of sources for S1 and S2 is 1,288 and 3,426, respectively. Splitting the RSG sample into two subgroups is necessary, since fitting all sources to all models without restrictions may lead to stars without observations in longer wavelengths being fitted to models with strong dust emission, even though these sources are supposedly without dust or only with very small amount of dust. If there is no restriction for S1, their DPR may be excessively overestimated, as shown in Figure~\ref{fig:fitre} (see also details below). Moreover, as a precaution, we also visually inspected all the SEDs and tentatively pre-fitted all models for all targets and discovered that, 26 RSGs in the S1 sample showed a significant rise in [3.6], [4.5], [5.8], and [8.0] bands (based on the visual inspection), these sources were then moved to the S2 sample.

Based on the best-fit results, we categorized all RSGs into three distinct types. The first type, named ``carbon-rich” (C-rich), is characterized by a significant continuous rise in the MIR region of the SED, without obvious emission peak. The second type, labeled as ``oxygen-rich" (O-rich), displays steep infrared emission peaks at 9.7 $\mu m$ and 18 $\mu m$. If a RSG showed no observable dust emission ($\tau_{9.7/11.3} = 0.0001$, which was the lower limit of optical depth in our model), regardless of whether it was fitted to the C-rich model or the O-rich model, we considered it as ``optically thin" (opt-thin). These optically thin RSG stars have SEDs that are similar to the central star, indicating that they have negligible amount of circumstellar dust and mass loss rate. In such cases, there was no difference between the C-rich model and the O-rich model, so we used the O-rich model for all optically thin RSGs.  During the fitting process, it was difficult to distinguish the chemical type of dust when the optical depth is small (e.g., $\tau_{9.7 / 11.3} \leq 0.005$). Therefore, we classified all sources with an optical depth less than 0.005 as O-rich, since most of the RSGs are expected to be O-rich. Such weak dust features, regardless of their type, are unlikely to have a significant impact on the calculation of the total DPR. Finally, we visually checked the fitting results of all RSGs and manually modified the fitting results of 141 sources, for which most of these sources had ambiguous dust properties that made it difficult to distinguish their types (we made the judgments based on our experience). 

\section{Result and Discussion}
\label{sec:result}
\begin{figure*}
    \centering
	\includegraphics[width=0.95\linewidth]{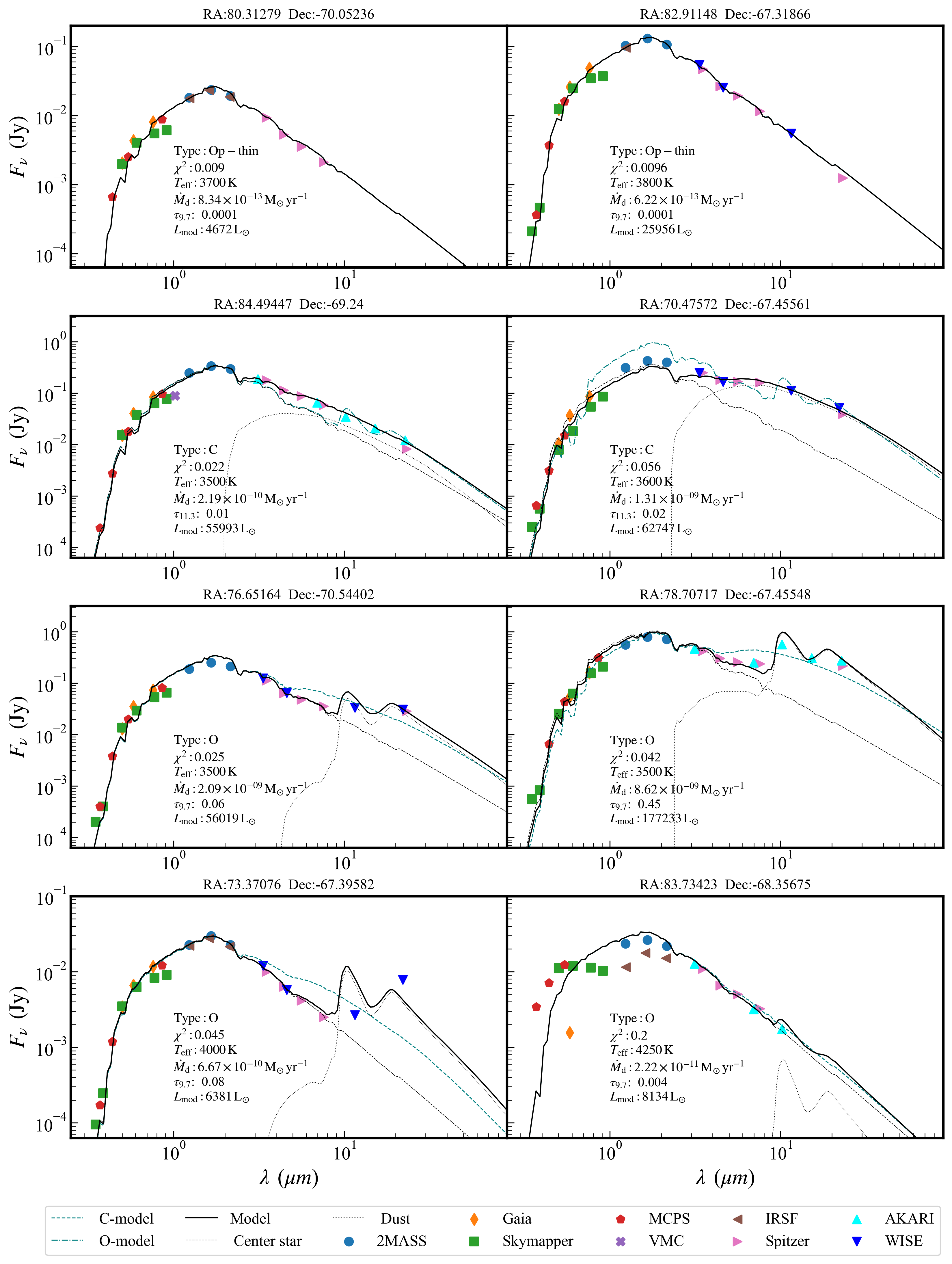}
    \caption{SED fits of eight example RSGs, the first, second and third row show the good SED fits for optically thin, C-rich and O-rich stars, respectively ($sflag=0$). The bottom row shows anomalous SED fits of two example stars, due to bad $W3/W4$ observation (left, $sflag=3$), and binarity (right, $sflag=2$)). }
    \label{fig:fitrsg}
\end{figure*}

Table~\ref{tab:result1} shows an example of the catalogue with photometric data, chemical composition, optical depth, luminosity, $T_{\rm{eff}}$, derived DPR, $\chi^2$ etc. Figure~\ref{fig:fitrsg} shows several examples of the fittings of optically thin, C-rich, and O-rich RSGs. The model SEDs generally match well with the measurements, with a few ones show poor fittings. The bottom row in Figure~\ref{fig:fitrsg} shows the SED of two examples of poorly fitted RSGs, which could be due to bad $W3/W4$ observations (left) or binarity (right). We defined a ``$sflag$," and for those normal RSGs, $sflag=0$. There are 55 sources identified as binaries due to their blue end flux excess, among them, the SEDs of 23 sources exhibit significant anomalies, and we set their $sflag$ to 2, for the remaining 32 sources whose fitting results were not affected, their $sflag$ was set to 2.5. Targets similar to the left panel in the bottom row, which have bad $W3$ and $W4$ observation, their $sflag$ was set to 3, we identified 245 such sources, which also exhibited anomalies in the luminosity-MLR diagram (see below). 

\movetabledown=2.5in
\begin{rotatetable}
\begin{deluxetable}{CCCCCCCCCCCCCCCCCCCCCCC}
\tabletypesize{\tiny}
\tablewidth{0pt} 
\tablecaption{\label{tab:result1} Catalog of the stellar and dust parameters for RSGs in LMC}
\tablehead{\colhead{
2MASS\_ID} &\colhead{RA} & \colhead{DEC} & \colhead{$E(B-V)$ $^{\rm a}$}& \colhead{$A_{\rm V}$}& \colhead{uSkymapper} & \colhead{e\_uSkymapper}& \colhead{......$^{\rm b}$}& \colhead{fluxspitzer24 }& \colhead{e\_fluxspitzer24 }& \colhead{sflag$^{\rm c}$} &\colhead{ $T_{\rm {eff}}$ }& \colhead{$\tau$ }& \colhead{mType} & \colhead{$\log{L_{\rm mod}}$} &\colhead{$\log{\dot{M}_{\rm d}} $} & \colhead{......$^{\rm d}$} & \colhead{fitindex $^{\rm e}$}& \colhead{$\chi^2$}\\     
\colhead { }&\colhead{(deg)}&\colhead{(deg)}&\colhead{(mag)}&\colhead{(mag)}&\colhead{(mag)}&\colhead{(mag)}&\colhead{ }&\colhead{(Jy)}&\colhead{(Jy)}&\colhead{ }&\colhead{(K)}&\colhead{ }&\colhead{ }&\colhead{($\rm L_{\sun}$)}&\colhead{($ \rm{M_{\sun} \, yr^{-1}}$)}&\colhead{ }&\colhead{ }&\colhead{ }}
\startdata
        05265311-6850002 & 81.72133 & -68.83341 & 0.165 & 0.563 & - & - & ....... & 6.79E-01 & 3.76E-03 & 0 & 4000 & 0.75 & O & 5.481  & -7.697  & ....... & 8441 & 0.0678  \\ 
        05384848-6905325 & 84.702 & -69.09238 & 0.029 & 0.098 & - & - & ....... & - & - & 0 & 3600 & 1.5 & O & 4.799  & -7.759  & ....... & 66974 & 0.0896  \\ 
        04533087-6917496 & 73.37865 & -69.29713 & 0.191 & 0.652 & - & - & ....... & - & - & 0 & 3700 & 1.5 & O & 4.845  & -7.790  & ....... & 7239 & 0.0262  \\ 
        05041413-6716143 & 76.0589 & -67.27065 & 0.02 & 0.067 & - & - & ....... & 2.85E-01 & 1.84E-03 & 0 & 3700 & 0.75 & O & 5.345  & -7.810  & ....... & 70993 & 0.0703  \\ 
        05374509-6920485 & 84.4379 & -69.34683 & 0.076 & 0.261 & - & - & ....... & 2.38E-01 & 1.98E-03 & 0 & 4250 & 0.95 & O & 5.087  & -7.812  & ....... & 66871 & 0.1361  \\ 
        05405919-6918361 & 85.24666 & -69.31005 & 0.134 & 0.457 & - & - & ....... & 2.40E-01 & 1.55E-03 & 0 & 3300 & 0.65 & O & 5.146  & -7.873  & ....... & 67989 & 0.0991  \\ 
        05355522-6909594 & 83.98012 & -69.16653 & 0.074 & 0.252 & - & - & ....... & 2.33E-01 & 1.50E-03 & 0 & 3300 & 0.65 & O & 5.146  & -7.873  & ....... & 67989 & 0.0962  \\ 
        05223109-6934051 & 80.62957 & -69.5681 & 0.034 & 0.115 & - & - & ....... & 1.13E-02 & 1.89E-04 & 0 & 3500 & 0.002 & C & 4.748  & -9.660  & ....... & 63986 & 0.0299  \\ 
        05352175-6913397 & 83.84065 & -69.22771 & 0.177 & 0.603 & - & - & ....... & - & - & 0 & 3400 & 0.0001 & Opt & 4.698  & -11.237  & ....... & 68523 & 0.0887  \\ 
        ...... & ...... & ...... & ...... & ...... & ...... & ...... & ...... & ...... & ...... & ...... & ...... & ...... & ...... & ...... & ...... & ...... & ...... & ...... \\
\enddata
\tablecomments{This table is published in its entirety in the machine-readable format. A portion is shown here for guidance regarding its form and content. \\$^{\rm a}$ The $E(B-V)$ value comes from the extinction map of  \citet{2022MNRAS.511.1317C}.\\ $^{\rm b}$ The complete table contains initial photometric data and error in all bands, covering a range of 0.3 to 24\,$\mu m$, and the complete table contains flux after extinction correction in all bands too. \\ $^{\rm c}$ Our defined classification flag, the sources with $sflag=0$ are normal sources, $sflag=2$ represents binaries or contamination, while $sflag=2.5$ indicates only slight flux excess that does not affect the fitting results. $sflag=3$ represents sources with bad W3/W4 band observational data.\\ $^{\rm d}$ The complete table includes more stellar and dust parameters, including total mass of the dust shell, inner radius temperature, mass loss timescale, etc. \\$^{\rm e}$ The best-fitting model for this RSG corresponds to a specific index number in the model library (referred to as ``index" in Table~\ref{tab:model2}). }
\end{deluxetable}
\end{rotatetable}

\begin{figure}
\centering
	\includegraphics[width=0.6\linewidth]{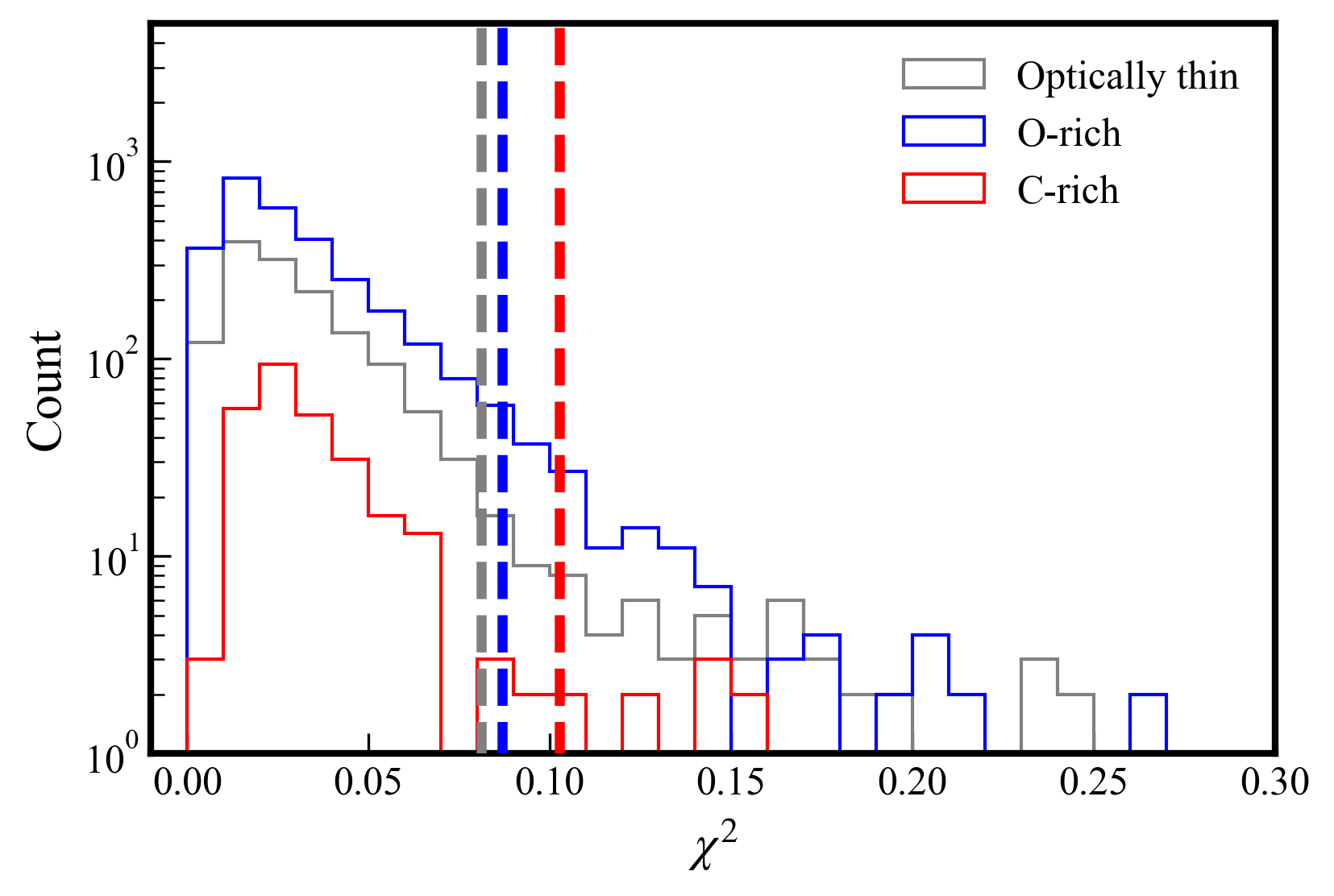}
    \caption{Histograms of the resultant $\chi^{2}$. The red, blue and gray bars represent those of the C-rich, O-rich and optically thin stars, respectively. The vertical lines show the 95th percentile $\chi^{2}$ values of the corresponding samples.}
    \label{fig:chidis}
\end{figure}

\subsection{Different types of the RSG dust shells}
The classification statistics of all the sample RSGs are presented in Table ~\ref{tab:DPRlmcrsg}. As a result, 1,444 RSGs  are classified as optically thin, while over half  are O-rich with 2,984 RSGs falling under this category. Only 286 RSGs  are classified as C-rich. Figure~\ref{fig:chidis} shows the $\chi^{2}$ distribution. The optically thin group appears to be better than O-rich and C-rich types, as indicated by the smaller $\chi^{2}$ values. We noted that in our study, we adopted a strict definition of ``optically thin”, where the optical depth of the model SED was $\tau_{\rm{9.7/11.3}} \leq 0.001$, and the infrared dust emission was minimal. If we relax this definition to include an optical depth $\tau_{\rm{9.7/11.3}} \leq 0.001$, the proportion of optically thin RSGs would increase to 35.41\%.

\begin{deluxetable*}{cccccccccc}
\tabletypesize{\footnotesize}
\tablecaption{\label{tab:DPRlmcrsg} Dust-production rate ($\dot{M}_{\rm d}$) of RSGs in the LMC}
\tablehead{\colhead{Type} & \colhead{N} & \colhead{Mean} &\colhead{ Median} & \colhead{Max}  & \colhead{Min}  & \colhead{Sum} &  \colhead{$\dot{M}_{\rm d}$\%$^a$} &  \colhead{$N\%$$^{b}$}\\
\colhead{ }& \colhead{}  & \colhead{ ($ \rm{M_{\sun}\,{yr^{-1}}}$)}& \colhead{($\rm{ M_{\sun}\,yr^{-1}}$)} & \colhead{($\rm{ M_{\sun}\,yr^{-1}}$)}&\colhead{ ($\rm{ M_{\sun}\,yr^{-1}}$) }& \colhead{($\rm{ M_{\sun}\,yr^{-1}}$)}&\colhead{ } & \colhead{ } }
\startdata
C-rich & 286 & $1.07\times 10^{-10}$ & $6.81\times 10^{-11}$ & $1.31\times 10^{-9}$ & $1.44\times 10^{-11}$ & $3.06\times 10^{-8}$ & 2.69\% & 6.07\%  \\ 
O-rich & 2,984 & $3.71\times 10^{-10}$ & $3.73\times 10^{-11}$ & $3.01\times 10^{-8}$ & $7.71\times 10^{-13}$ & $1.11\times 10^{-6}$ & 97.24\% & 63.30\%  \\ 
Optically thin & 1,444 & $5.81\times 10^{-13}$ & $5.90\times 10^{-13}$ & $5.79\times 10^{-12}$ & $1.97\times 10^{-13}$ & $8.39\times 10^{-10}$ & 0.07\% & 30.63\%  \\
All & 4,714 & $2.41\times 10^{-10}$ & $2.48\times 10^{-11}$ & $3.01\times 10^{-8}$ & $1.97\times 10^{-13}$ & $1.14\times 10^{-6}$ & 100.00\% & 100.00\%  \\ 
All($sflag\neq 3$ $^{\rm c}$) & 4,469 & $2.35\times 10^{-10}$ & $2.36\times 10^{-11}$ & $3.01\times 10^{-8}$ & $1.97\times 10^{-13}$ &$1.05\times 10^{-6}$ & 92.18\% & 94.80\%  \\
All( $sflag=0$ $^{\rm d}$) & 4,414 & $2.27\times 10^{-10}$ & $2.36\times 10^{-11}$ & $3.01\times 10^{-8}$ & $1.97\times 10^{-13}$ &$1.00\times 10^{-6}$ & 88.22\% & 93.64\%  \\
\enddata
\tablecomments{$^{\rm a}$ Proportion to total DPR. $^{\rm b}$ Proportion to total number of RSG. $^{\rm c}$ $sflag=3$ represents sources with bad W3/W4 band observational data. $^{\rm d}$$sflag=0$ represents sources that are considered normal.}
\end{deluxetable*}

\begin{figure*}
	\includegraphics[width=0.95\linewidth]{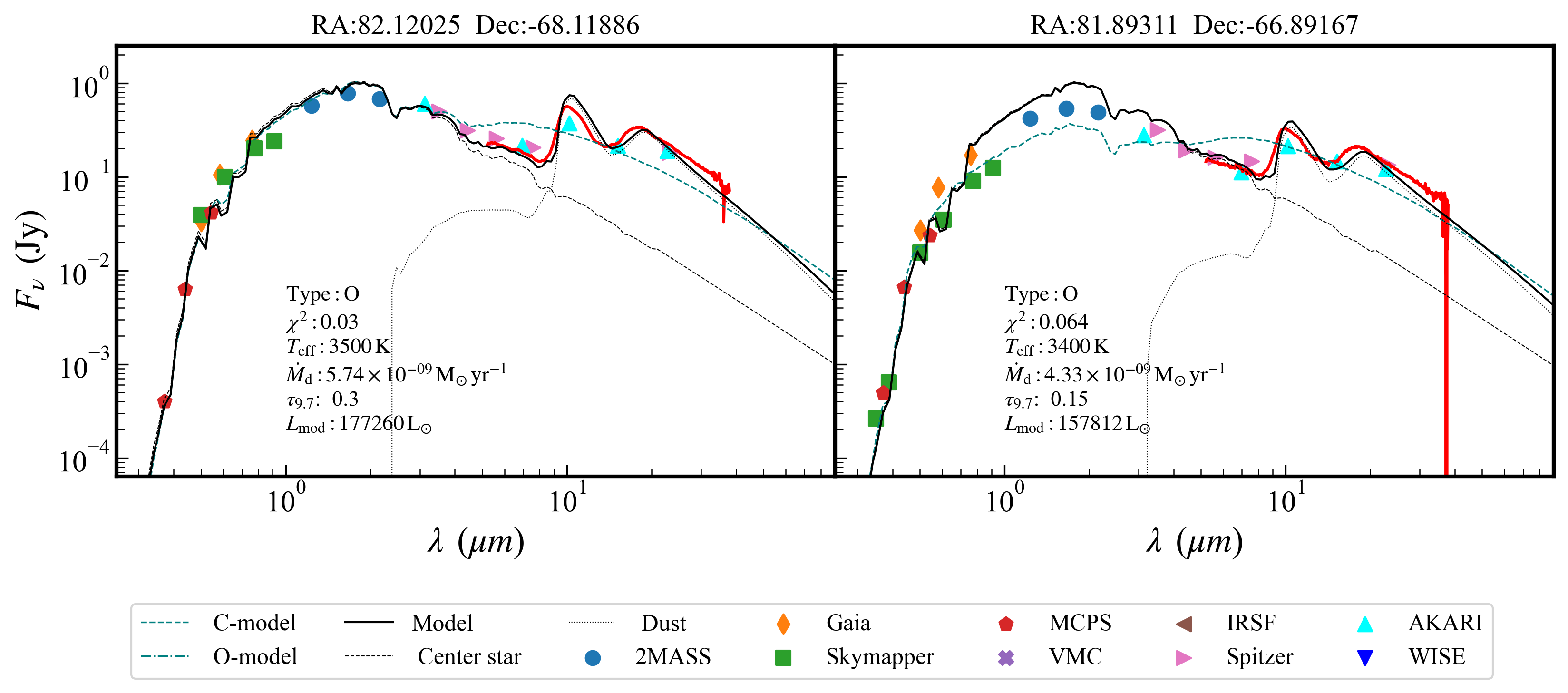}
    \caption{Examples of IRS spectra with high DPR and their classification in our sample. The red line represents the IRS spectral data, and other marks are the same as Figure~\ref{fig:fitrsg}.}
    \label{fig:fitrsg2}
\end{figure*}

To verify the accuracy of our classification, we crossmatched our catalogue with the spectroscopic data from the Spitzer IRS (InfraRed Spectrograph) Enhanced Products \citep{2004ApJS..154...18H} using a search radius of $1^{\arcsec}$. We identified 50 sources that had matches.  Visual inspection of their spectra revealed that 43 of the 50 sources were correctly classified, resulting in an accuracy of 86\%. The 7 misclassified sources were all O-rich stars mistakenly identified as C-rich. This is possibly due to that there are both silicate emission peaks and prominent continuous emission in the NIR to MIR bands in their SEDs. We corrected the classification of these sources in our resultant catalogue based on their spectra. Figure~\ref{fig:fitrsg2} shows the examples of IRS spectra with high DPR and their classification in our sample. 

The classification of lower mass stars such as AGB stars as C-rich can be attributed to carbon being brought to the surface of the stars through deep convection and dredge-up processes \citep{2005ARA&A..43..435H,2008A&A...482..883M,2019MNRAS.485.5666P}. However, the formation mechanism of carbon-rich RSGs, which lacks the mechanism to bring carbon to the surface due to their higher mass, requires further investigation. 

\begin{figure}
	\includegraphics[width=\linewidth]{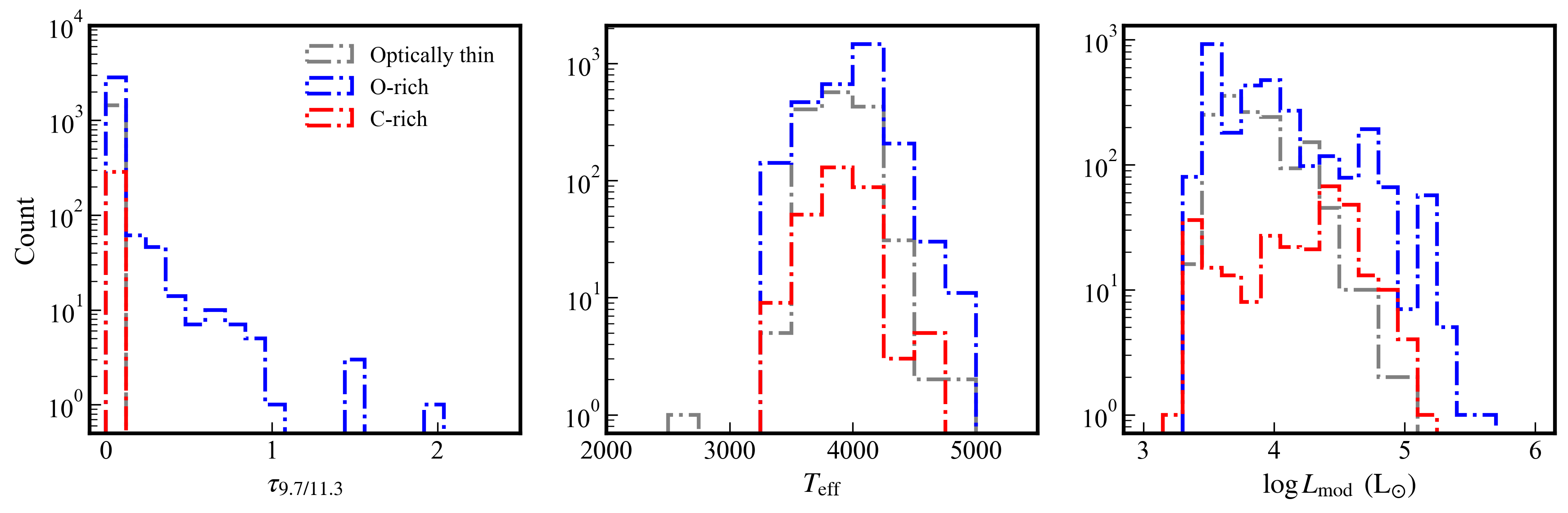}
    \caption{Histograms of the best-fit values of optical depth (left), effective temperature (middle) and luminosity (right). Red, blue, and gray bars represent those of the C-rich, O-rich and optically thin RSGs, respectively.}
    \label{fig:tefflumtau}
\end{figure}

The distributions of the resultant values of optical depth, effective temperature, and luminosity of the RSGs with different types are shown in Figure~\ref{fig:tefflumtau}. Our results indicate that almost all RSGs (96.80\%) have temperatures between 3,300\,K and 4,500\,K, and luminosities between 3,000 and 400,000\,$\rm{L_{\sun}}$. This is consistent with previous findings \citep{1998ApJ...505..793M,2003AJ....126.2867M,2010NewAR..54....1L,2017ars..book.....L}, which suggests the purity of our sample. No significant differences are visible in the distributions of these parameters among the different types, except for that the C-rich RSGs tend to have smaller optical depths.

\subsection{The resultant DPR and MLR}
\label{subsec:result2}
\begin{figure*}
        \centering
	\includegraphics[width=0.9\linewidth]{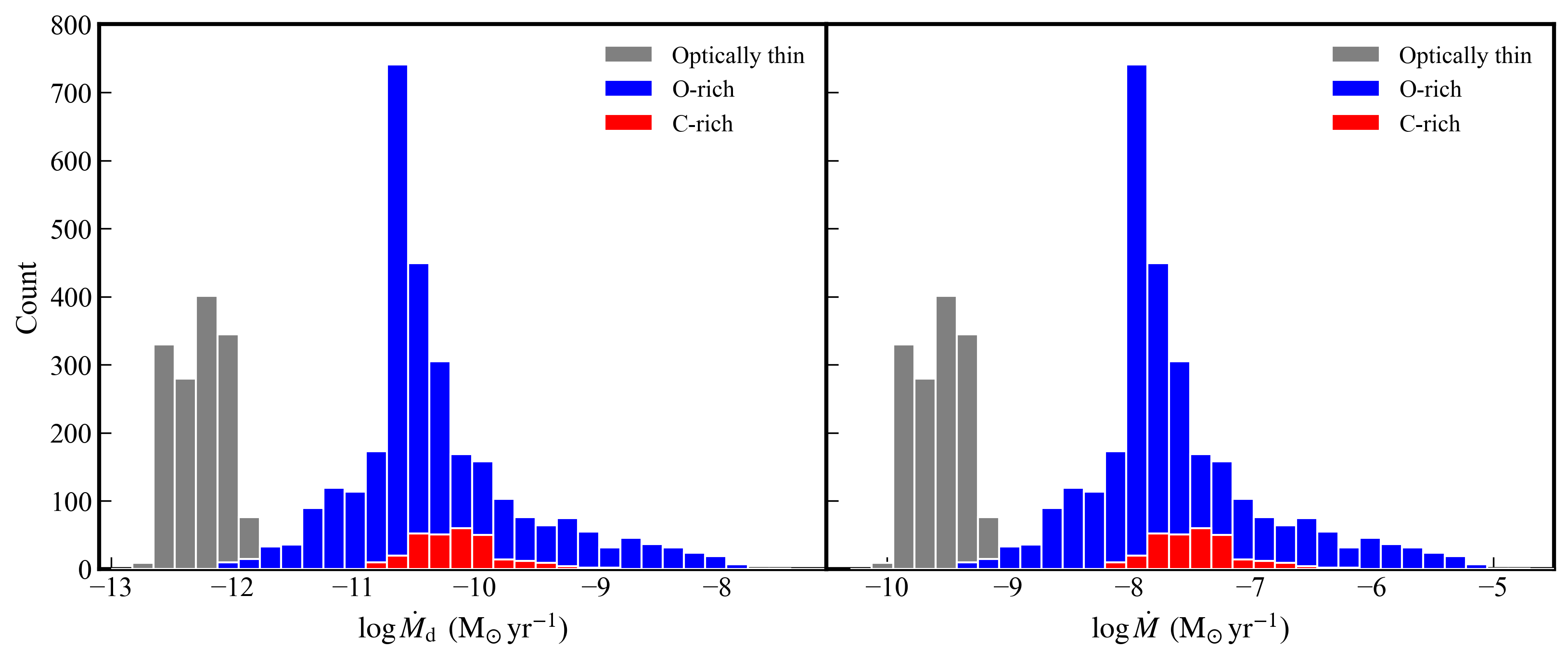}
    \caption{Histograms of the best-fit values of DPR ($\dot{M_{\rm d}}$, left) and MLR ($\dot{M}$, right). Red, blue, and gray bars represent those of the C-rich, O-rich and optically thin RSGs, respectively.}
    \label{fig:lmcrsgmlr}
\end{figure*}

We assumed a gas-to-dust ratio of $\phi = 500$ \citep{2012ApJ...753...71R} for the LMC. The MLR value of each RSG was calculated based on its best-fit DPR value and the assumed gas-to-dust ratio. The histograms of derived DPR and MLR for the RSG sample are presented in Figure~\ref{fig:lmcrsgmlr}. Optically thin RSGs have extremely low MLR. The C-rich RSGs exhibit a higher MLR than the O-rich ones. This could be partly due to our classification method which only recognizes RSGs with obvious carbon emission.

The typical MLR value of our sample RSGs is $10^{-8}$ to $10^{-7}\, \rm{M_{\sun}\, yr^{-1}}$, while a few can reach to $10^{-5}\, \rm{M_{\sun}\, yr^{-1}}$. Table~\ref{tab:DPRlmcrsg} provides detailed DPR values for different types of RSGs. The total DPR including is $1.14 \times 10^{-6}\, \rm{M_{\sun}\, yr^{-1}}$, with O-rich RSGs contributing the most (97.24\%), followed by C-rich RSGs and optically thin RSGs. The average DPR of our sample RSGs is about $2.41\times10^{-10}\, \rm{M_{\sun}\, yr^{-1}}$, with a median value of about $2.48\times 10^{-11}\,\rm{M_{\sun}\, yr^{-1}}$. 208 RSGs with a DPR higher than $10^{-9}\,\rm{M_{\sun}\, yr^{-1}}$ contributed 76.57\% of the total DPR. The fitting SEDs for several RSGs with high DPR are shown in Figure~\ref{fig:fitrsg2}.

\begin{figure*}
	\includegraphics[width=1\linewidth]{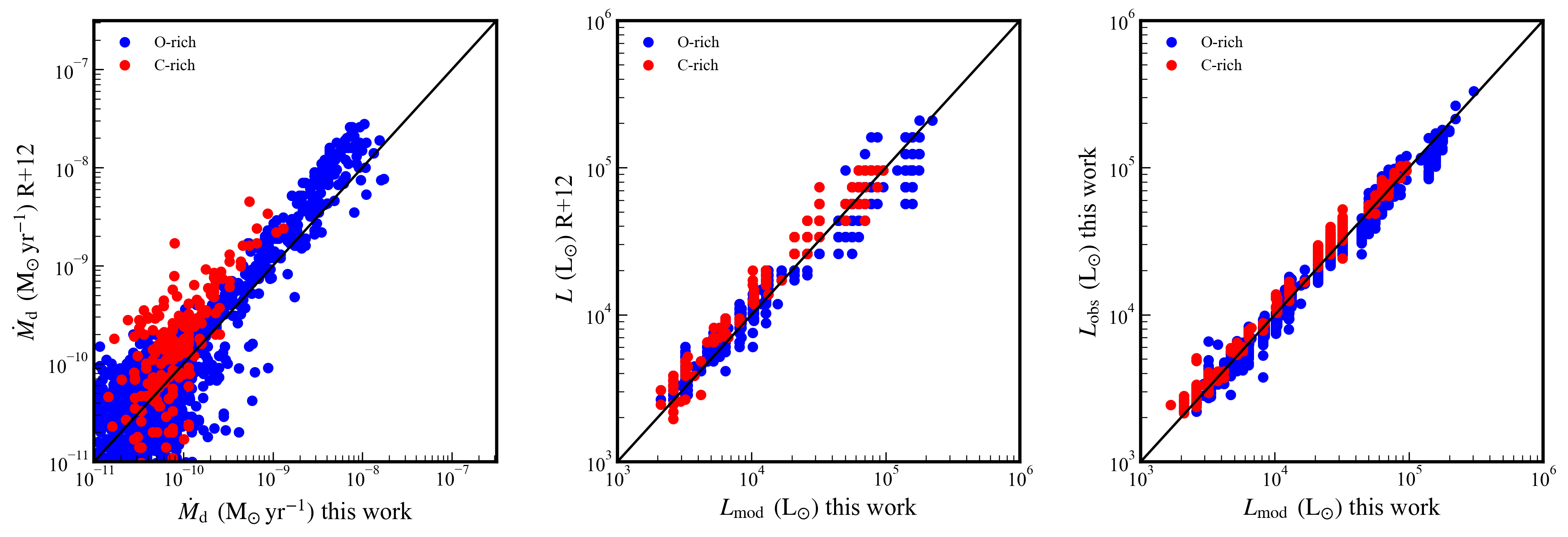}
    \caption{Comparison between our DPR and luminosity and the results of R+12 (left, middle), and the panel on the right is the comparison between the luminosity obtained by integrating the observed SED and the luminosity obtained by integrating the best fit model SED.}
    \label{fig:vs}
\end{figure*}

The values of MLR and DPR of RSGs in the LMC were also estimated in previous work. R+12 obtained the global DPR of the LMC from both RSGs and AGB stars, which was around $2.1 \times 10^{-5}\, \rm{M_{\sun}\, yr^{-1}}$. Among them, RSGs accounted for 9.4\%, and the total DPR of all RSGs was about $2.0 \times 10^{-6}\, \rm{M_{\sun}\, yr^{-1}}$. B+12 found that the total DPR of 3,908 RSGs in the LMC was about $2.4 \times 10^{-7}\, \rm{M_{\sun}\, yr^{-1}}$ with an average DPR of about $6.1 \times 10^{-11}\, \rm{M_{\sun}\, yr^{-1}}$. S+16 estimated that the total DPR of 1,410 SMC RSGs in their sample was $4.6 \times 10^{-8}\, \rm{M_{\sun}\, yr^{-1}}$, with an average DPR of $3.3 \times 10^{-11}\, \rm{M_{\sun}\, yr^{-1}}$. Our results are basically agreement with B+12, R+12 and S+16 within the uncertainties. Table~\ref{tab:DPRlmcrsgvs} shows our results and the comparison with previous works. Compared to the result of R+12 (all evolved star sammple, including AGB and RSGs), our DPR (only RSGs with $sflag=0$ are included) is slightly lower as shown in Figure~\ref{fig:vs}. But our resultant luminosity values are consistent with theirs. To study the impact of sample size differences on the results, we cross-matched our sample with their RSG sample and identified 4,213 common sources. For these sources, the cumulative R+12 DPR was $1.45 \times 10^{-6}\, \rm{M_{\sun}\, yr^{-1}}$, which is a 27.5\% reduction compared to the result of their original result. Sample size undoubtedly directly affects the results. For the 4,213 RSGs, our calculated cumulative DPR is $9.0 \times 10^{-7}\, \rm{M_{\sun}\, yr^{-1}}$, which is still slightly lower than that of R+12. This overall lower trend may be attributed to differences in model settings, fitting strategies, or other factors (see the detailed disscusion in Section 5.1 of Y+23).

Our total DPR is 4.6 times that of B+12, we cautiously regard this difference as within an acceptable range, particularly considering the difference in our research approaches. B+12 employed a hybrid approach in which they initially fitted the DPR for several evolved stars, then established the relationship between DPR and [8.0] excess, finally applied this relationship to the entire sample. The sample they utilized was from \citet{2011AJ....142..103B}, for which the sample was selected based on 2MASS \citep{2006AJ....131.1163S} and SAGE \citep{2010AJ....140.1868W}, the spatial coverage of the sample is smaller than that of \citet{2021ApJ...923..232R}. We identified 3,855 sources in common. For those sources, our calculated total DPR is $6.74 \times 10^{-7}\, \rm{M_{\sun}\, yr^{-1}}$.

\begin{deluxetable}{CCCCC}
\tablecaption{\label{tab:DPRlmcrsgvs} Comparison of total DPRs between different works}
\tablehead{\colhead{  }&\colhead{Galaxy}&\colhead{N}& \colhead{Mean}  & \colhead{Sum}  \\
\colhead{}& \colhead{}&\colhead{}&  \colhead{($ \rm{M_{\sun}\,{yr^{-1}}}$)}& \colhead{($\rm{ M_{\sun}\,yr^{-1}}$)}}
\startdata
\rm{This\ work} &\rm{LMC}&4,714 & $2.4\times 10^{-10}$  & $1.1\times 10^{-6}$  \\ 
\rm{Riebel\ et\ al.\ (2012)} &\rm{LMC}& 5,876&$3.4\times 
10^{-10}$&$2.0\times 10^{-6}$\\
\rm{Boyer\ et\ al.\ (2012)}  &\rm{LMC}& 3,908&$6.1\times 10^{-11}$&$2.4\times 10^{-7}$\\
\rm{Boyer\ et\ al.\ (2012)} &\rm{SMC}& 2,611&$1.2\times 10^{-11}$&$3.1\times 10^{-8}$\\
\rm{Srinivasan \ et\ al.\ (2016)} &\rm{SMC}& 1,410&$3.3\times 10^{-11}$&$4.6\times 10^{-8}$\\
\enddata
\end{deluxetable}

The average MLR of RSGs in M31 and M33 was found to be
about $2.0 \times 10^{-5}\, \rm{M_{\sun}\, yr^{-1}}$ by W+21. Y+23 found that the typical MLR of RSGs in the SMC is about $10^{-6}\, \rm{M_{\sun}\, yr^{-1}}$. These values are larger than our results. The determination of MLR is particularly sensitive to the choice of parameters such as the optical constants, gas-to-dust ratio, and dust shell expansion speed, etc., which can cause the difference in results to be up to one order of magnitude \citep{2016MNRAS.457.2814S, 2023A&A...676A..84Y}. \citet[hereafter GS18]{2018A&A...609A.114G} fitted the SEDs and IRS spectra of nearly 400 evolved stars in the LMC, SMC, and other dwarf spheroidal galaxies in the Local Group. They estimated the MLR for each target ranging from $10^{-4}$ to $10^{-11}\, \rm{M_{\sun}\, yr^{-1}}$, which also highly depended on the adopted optical constants.

Moreover, we varied some parameters and methods to assess their impact on our DPR and MLR results. The statistical results are shown in Table~\ref{tab:effect}. When we used the optical constant of \citet[hereafter Do+95]{1995A&A...300..503D} to generate the model for fitting, the resulted DPR was 22\% lower and the chemical classification were also changed. The number of optically thin sources increased to 1,547, while the number of C-rich sources increased to 300, and the number of O-rich sources decreased to 2,867. Their contributions to total DPR were also changed, for which the detailed statistical results are listed in Table~\ref{tab:DPRdo}.

{\catcode`\&=11
\gdef\Dooo {\cite{1995A&A...300..503D}}}

\begin{deluxetable}{ccccc}
\tablecaption{\label{tab:effect} Effect of different parameter settings on DPR}
\tablehead{\colhead{ }& \colhead{ Mean} & \colhead{Median} & \colhead{Sum}&  \colhead{$d\dot{M}_{\rm{d}}\%^{a}$}\\
\colhead{ }&  \colhead{ ($ \rm{M_{\sun}\,{yr^{-1}}}$)}& \colhead{($ \rm{M_{\sun}\,{yr^{-1}}}$)} & \colhead{($ \rm{M_{\sun}\,{yr^{-1}}}$) }& \colhead{}}   
\startdata
This work & $2.41\times 10^{-10}$ & $2.48\times 10^{-11}$ &$1.14\times 10^{-6}$ & - \\
Do+95$^{\rm b}$ & $1.88\times 10^{-10}$ & $1.95\times 10^{-11}$ & $8.88\times 10^{-7}$ & $-22\%$ \\ 
$E(B-V)=0.1$ and $R_{\rm V}=3.1$ & $2.41\times 10^{-10}$ & $2.48\times 10^{-11}$ &$1.14\times 10^{-6}$& 0.00\% \\ 
$\chi_{m}^{2}$(Eq3)$^{\rm c}$ & $1.89\times 10^{-10}$ & $2.78\times 10^{-11}$ &$8.92\times 10^{-7}$&$-22\%$\\ 
scaling $v_{\rm{exp}}$$^{\rm d}$& $1.84\times 10^{-10}$ & $9.94\times 10^{-12}$ &$8.69\times 10^{-7}$&$-24\%$\\
$v_{\rm{exp}}= 25\,\rm{km\,s^{-1}}$ & $6.03\times 10^{-10}$ & $6.20\times 10^{-11}$ &$2.84\times 10^{-6}$&$+150\%$ \\ 
\enddata
\tablecomments{$^{\rm a}$ The change rate of DPR compared to the original parameter setting. $^{\rm b}$The fitting results using the optical constants obtained from \Dooo.  $^{\rm c}$As defined in Equation 3 of Y+23. $^{\rm d}$ The result obtained by scaling the wind speed using Equation~\ref{eq:vexp}.}
\end{deluxetable}

\begin{deluxetable*}{ccccccccc}
\tablecaption{\label{tab:DPRdo}Dust-production rate ($\dot{M}_{\rm d}$) of RSGs in the LMC using Do+95 optiacl constans}

\tablehead{\colhead{Type} & \colhead{N} & \colhead{Mean} &\colhead{ Median} & \colhead{Max}  & \colhead{Min}  & \colhead{Sum} &  \colhead{$\dot{M}_{\rm d}$\%$^{\rm a}$} &  \colhead{$N\%$$^{\rm b}$}\\
\colhead{ }& \colhead{}  & \colhead{ ($ \rm{M_{\sun}\,{yr^{-1}}}$)}& \colhead{($\rm{ M_{\sun}\,yr^{-1}}$)} & \colhead{($\rm{ M_{\sun}\,yr^{-1}}$)}&\colhead{ ($\rm{ M_{\sun}\,yr^{-1}}$) }& \colhead{($\rm{ M_{\sun}\,yr^{-1}}$)}&\colhead{ } & \colhead{ } }
\startdata
C-rich & 300 & $2.23\times 10^{-10}$ & $7.57\times 10^{-11}$ & $4.92\times 10^{-9}$ & $1.44\times 10^{-11}$ & $6.69\times 10^{-8}$ & 7.54\% & 6.36\%  \\
O-rich & 2,867 & $2.86\times 10^{-10}$ & $3.90\times 10^{-11}$ & $1.60\times 10^{-8}$ & $1.82\times 10^{-12}$ & $8.19\times 10^{-7}$ & 92.27\% & 60.82\%  \\
Optically thin & 1,547 & $1.08\times 10^{-12}$ & $9.19\times 10^{-13}$ & $5.34\times 10^{-12}$ & $3.90\times 10^{-13}$ & $1.67\times 10^{-9}$ & 0.19\% & 32.82\% \\ 
All & 4,714 & $1.88\times 10^{-10}$ & $1.95\times 10^{-11}$ & $1.60\times 10^{-8}$ & $3.90\times 10^{-13}$ & $8.88\times 10^{-7}$ & 100.00\% & 100.00\% \\
All ($sflag\neq 3$ $^{\rm c}$) & 4,469 & $1.87\times 10^{-10}$ & $1.68\times 10^{-11}$ & $1.60\times 10^{-8}$ & $3.90\times 10^{-13}$ & $8.35\times 10^{-7}$ & 94.02\% & 94.80\% \\
All ( $sflag=0$ $^{\rm d}$) & 4,414 & $1.82\times 10^{-10}$ & $1.64\times 10^{-11}$ & $1.60\times 10^{-8}$ & $3.90\times 10^{-13}$ & $8.03\times 10^{-7}$ & 90.50\% & 93.64\% \\
All (DL84$^{\rm e}$)& 4,714 & $2.41\times 10^{-10}$ & $2.48\times 10^{-11}$ & $3.01\times 10^{-8}$ & $1.97\times 10^{-13}$ & $1.14\times 10^{-6}$ & 128\% & 100.00\%  
\enddata
\tablecomments{$^{\rm a}$ Proportion to total DPR. $^{\rm b}$ Proportion to total number of RSG. $^{\rm c}$ $sflag=3$ represents sources with bad W3/W4 band observational data. $^{\rm d}$$sflag=0$ represents sources that are considered normal. $^{\rm e}$The fitting results using the optical constants obtained from \citet{1984ApJ...285...89D}.}
\end{deluxetable*}

In order to explore the factors that affect the DPR results, We also changed the extinction correction, by adopting $E(B-V)=0.1$ and $R_{\rm V}=3.1$ \citep{2011ApJ...737..103S,2020ApJ...891...57F}, using the extinction coefficients of \citet{2019ApJ...877..116W}. The change in cumulative DPR is less than 0.01\%, only one target has changed its classifiction, indicating that different extinction correction methods we used has almost no effect on the derived DPR.We also employed Equation 3 of Y+23 for model fitting, for which the result showed a 22\% reduction in cumulative DPR. The definition of fitting criteria is also an nonignorable factor that affects the results.

In addition, we scaled the expansion speeds of the dust shell by using the \citet{2006ASPC..353..211V} scaling relation:

\begin{equation}
\label{eq:vexp}
    \frac{v_{\rm{exp}}}{10\,\rm{km\,s^{-1}}}=(\frac{L}{3\times 10^{4}\,\rm{L_{\sun}}})^{1/4} \times (\frac{\phi}{200})^{-1/2}
\end{equation}

Using Equation~\ref{eq:vexp} leads to a reduction in cumulative DPR, consistent with the findings of S+16. If we adopt $v_{\rm{exp}}= 25 \rm{\,km\,s^{-1}}$ as done by \citet{2022ApJ...933...41B}, our result would be 2.5 times higher than the original value. However, the expansion rate is highly uncertain and dependent on many parameters, like the properties of the central star, metallicity, dust grain size, etc., making it a significant source of DPR uncertainty.

For the optically thin sources, their DPR may be very low, but their uncertainty is also considerable. The derived DPR may not represent their true DPR, as they are only fitted to the lower limit of the template DPR, which can be as low as $ 10^{-13}\, \rm{M_{\sun}\, yr^{-1}}$, which may not be physically meaningful. Additionally, the light variation period
of RSGs, typically about 300 to 800 days \citep{2011ApJ...727...53Y,2019MNRAS.487.4832C,2019ApJS..241...35R}, which may also affect our results, as our photometric data may not cover a complete period. Moreover, photometric errors and data quality also cannot be ignored. See more disscusions in Y+23.

In summary, the impact factors of DPR obtained through the SED fitting method are complex. We can only provide a rough estimation, and the uncertainty of the DPR may reach up to one order of magnitude or even more. 

\subsection{The MLR and stellar parameters relations}
\label{subsec:3}
\begin{figure*}
	\includegraphics[width=\linewidth]{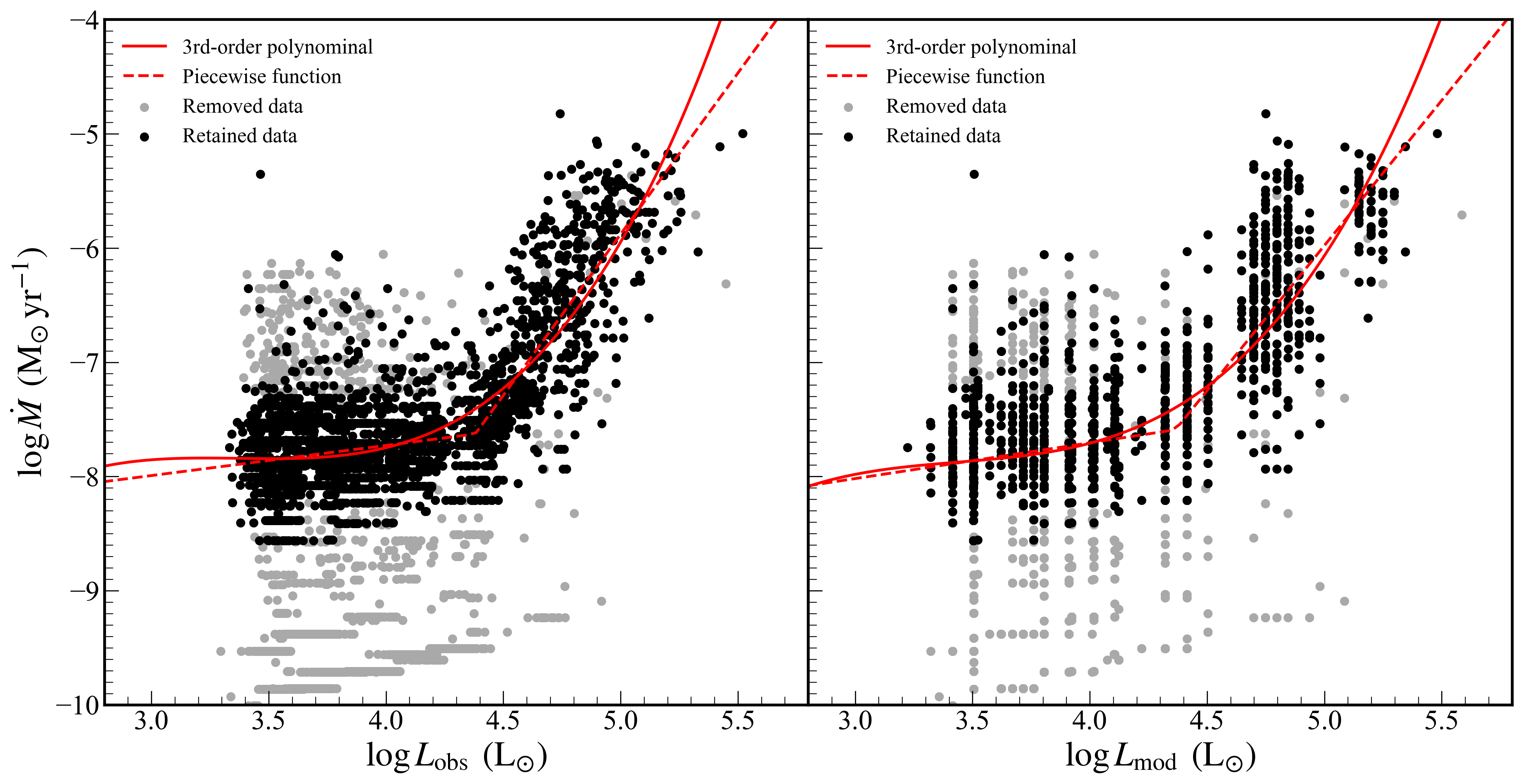}
    \caption{Luminosity-MLR diagram. The left panel shows the luminosity obtained by integrating the observed SED ($L_{\rm{obs}}$), and the right panel shows the luminosity obtained by integrating the best-fit model $L_{\rm{mod}}$. The excluded and retained stars are marked by gray and black, respectively. The red curves represent the best-fitted relations, which are labeled in the diagrams.}
    \label{fig:fitdata}
\end{figure*}

The relations between MLR and stellar parameters, such as the luminosity, temperature, radius, and initial mass, etc, are extensively calculated in the literature. In this work, we utilized a third-order polynomial and a piecewise function to fit the data and obtained the expressions for luminosity (both $L_{\rm{obs}}$ and $L_{\rm{mod}}$) and the MLR, respectively. Figure~\ref{fig:fitdata} depicts the relationship between luminosity and MLR. To ensure the accuracy of the fitting, optically thin targets, or targets have a high level of uncertainty ($\tau_{\rm{9.7 / 11.3}} \leq$ 0.001 and $sflag=2$, 2.5, 3) are excluded. These $sflag=3$ targets (245 stars) are mostly located in the region where $\log{(L_{\rm{obs}}/L_{\sun})}<4.5 $ and $\log{\dot{M}} \geq -7.5 $, characterized by low luminosity and high MLR, they are considered to be pseudo sources or affected by other nearby sources. The SED of an example star is shown in the bottom row left panel of Figure~\ref{fig:fitrsg}. These stars are excluded from the MLR-luminosity functions \footnote{Within the luminosity range of $\log{(L_{\rm{obs}}/L_{\sun})}= 3.5$ to 4, after removing the sources with $sflag=3$, there are still some sources with slightly higher MLR, $\log{\dot{M}}$ approximately ranging from $-7$ to$ -6$. These sources like those with $sflag=3$, exhibiting unusual $W3$ and $W4$ data, they also have data for $S11$ or [24], so they are not classified into the $sflag=3$ category. Some contamination from AGB stars is inevitable, particularly in the low-luminosity region. Overall, these few anomalous sources do not significantly impact our fitting results.}. As a result, we have third-order polynomial:

\begin{alignat}{4}   
\label{eq:3rdobs}
    \log{\dot{M}}&= 0.45\times [\log{(L_{\rm{obs}}/L_{\sun})}]^{3} &&- 4.55\times[\log{(L_{\rm{obs}}/L_{\sun})}]^{2} 
    &&+ 15.32\times[\log{(L_{\rm{obs}}/L_{\sun})}] &&- 25.00\\
    \log{\dot{M}}&= 0.41\times[\log{(L_{\rm{mod}}/L_{\sun})}]^{3} &&- 4.20\times[\log{(L_{\rm{mod}}/L_{\sun})}]^{2} 
    &&+ 14.61\times[\log{(L_{\rm{mod}}/L_{\sun})}] &&-25.00
\end{alignat}
piecewise function:
\begin{alignat}{3}
    \log{\dot{M}}&= 0.26\times[\log{(L_{obs}/L_{\sun})}] - 8.78\quad&&(\log{(L_{obs}/L_{\sun})} <&& 4.38)\notag\\
    &= 2.82\times[\log{(L_{obs}/L_{\sun})}] - 19.99\quad  &&(\log{(L_{obs}/L_{\sun})} \geq&& 4.38)
\end{alignat}
\begin{alignat}{3}
    \log{\dot{M}}&= 0.31\times[\log{(L_{mod}/L_{\sun})}] - 8.96\quad&&(\log{(L_{mod}/L_{\sun})} <&& 4.37)\notag\\
   &= 2.54\times[\log{(L_{mod}/L_{\sun})}] - 18.68\quad&&(\log{(L_{mod}/L_{\sun})} \geq&& 4.37)
\end{alignat}

\begin{figure*}
	\includegraphics[width=1\linewidth]{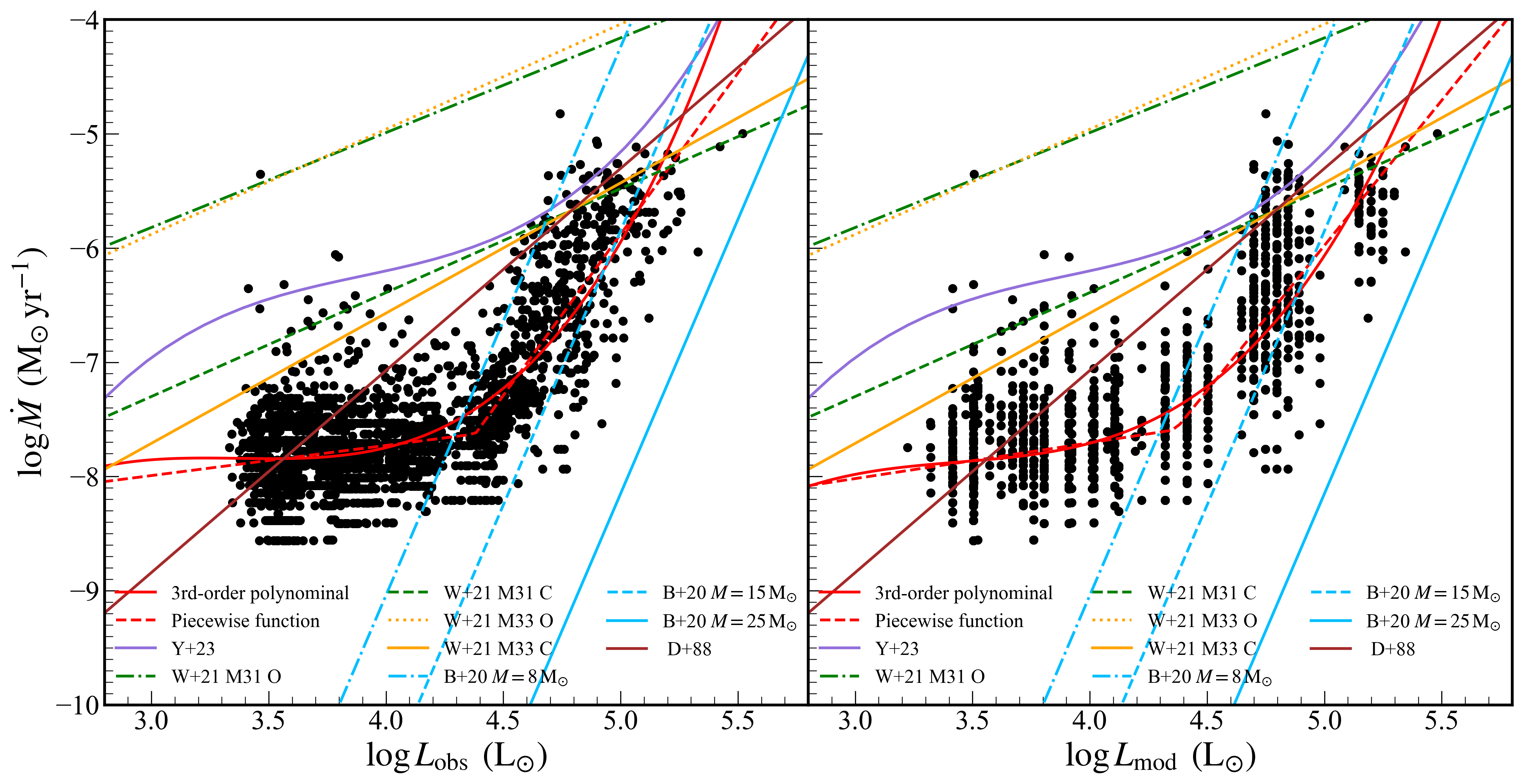}
    \caption{Comparisons of our luminosity-MLR relations with those from previous works. Left panel show the luminosity obtained by integrating the observation SED ($L_{\rm{obs}}$), and right panel shows the luminosity obtained by integrating the best fit model $L_{\rm{mod}}$. The retained stars are marked by black. Different curves are the relations from our work and the literature, which are labelled in the diagrams (see Section~\ref{subsec:3}).}
    \label{fig:others}
\end{figure*}

We compared our results with those from the literature, including \citet[hereafter D+88]{1988A&AS...72..259D}, \citet[hereafter B+20]{2020MNRAS.492.5994B}, W+21 and Y+23 as shown in Figure~\ref{fig:others}. The Figure shows that our resulted MLR values are slightly lower than those of D+88, W+21, and Y+23, particularly in the case of low luminosity. D+88 formulated a correlation from a relatively small sample, while both W+21 and Y+23 used the DUSTY radiation transfer model instead of 2-DUST with various parameter settings and different selections of optical constants, which might significantly affect the results. Our relations appear to be consistent with the results of B+20, especially for regions with $\log{(L_{\rm{obs}}/L_{\sun})} \geq 4.0 $. Our sample also shows similar turning point at $\log{(L_{\rm{obs}}/L_{\sun})} \approx 4.4 $ on the luminosity versus MLR diagram as reported by Y+23, which they call a ``knee-like" shape, at $\log{(L_{\rm{obs}}/L_{\sun})} \approx 4.6 $ in the SMC. The similar trend with different turning point of luminosity-MLR relation may be due to the difference in metallicity between the LMC and SMC.

Recently, \citet{2022ApJ...933...41B} re-evaluated the MLR of Dust-enshrouded RSGs  and found that the relations of \citet{2005A&A...438..273V} likely overestimated the MLR of RSGs. B+20 also showed that the RSGs with smaller MLRs might not be able to lose all of their dust shells, returned to the blue end in the H-R diagram and became hydrogen-poor 
supernova. Similarly, our results indicate that the RSGs can not lose several solar masses with static stellar wind. Unlike AGB stars, RSGs may not be the main contributors of dust in the interstellar medium.

\begin{figure*}
	\includegraphics[width=\linewidth]{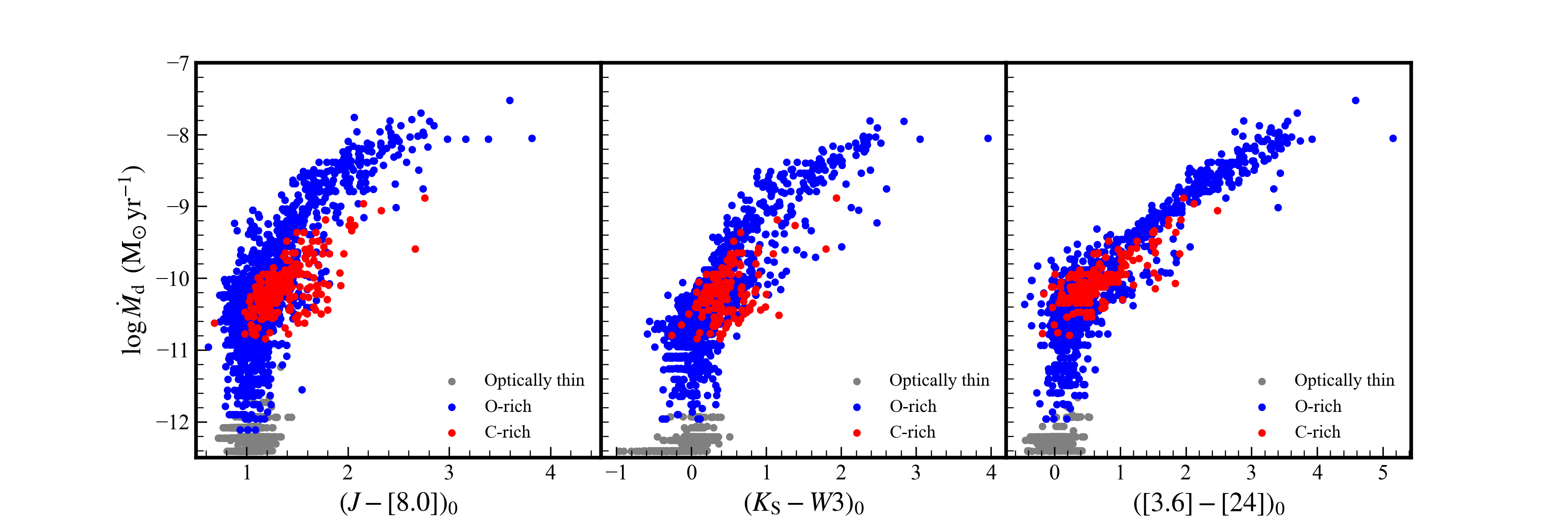}
    \caption{Stellar colors  versus DPR diagrams for the optically thin (gray dots), O-rich (blue dots) and C-rich (red dots) RSG sample.}
    \label{fig:cmlr}
\end{figure*}

\begin{figure}
\centering
	\includegraphics[width=0.6\linewidth]{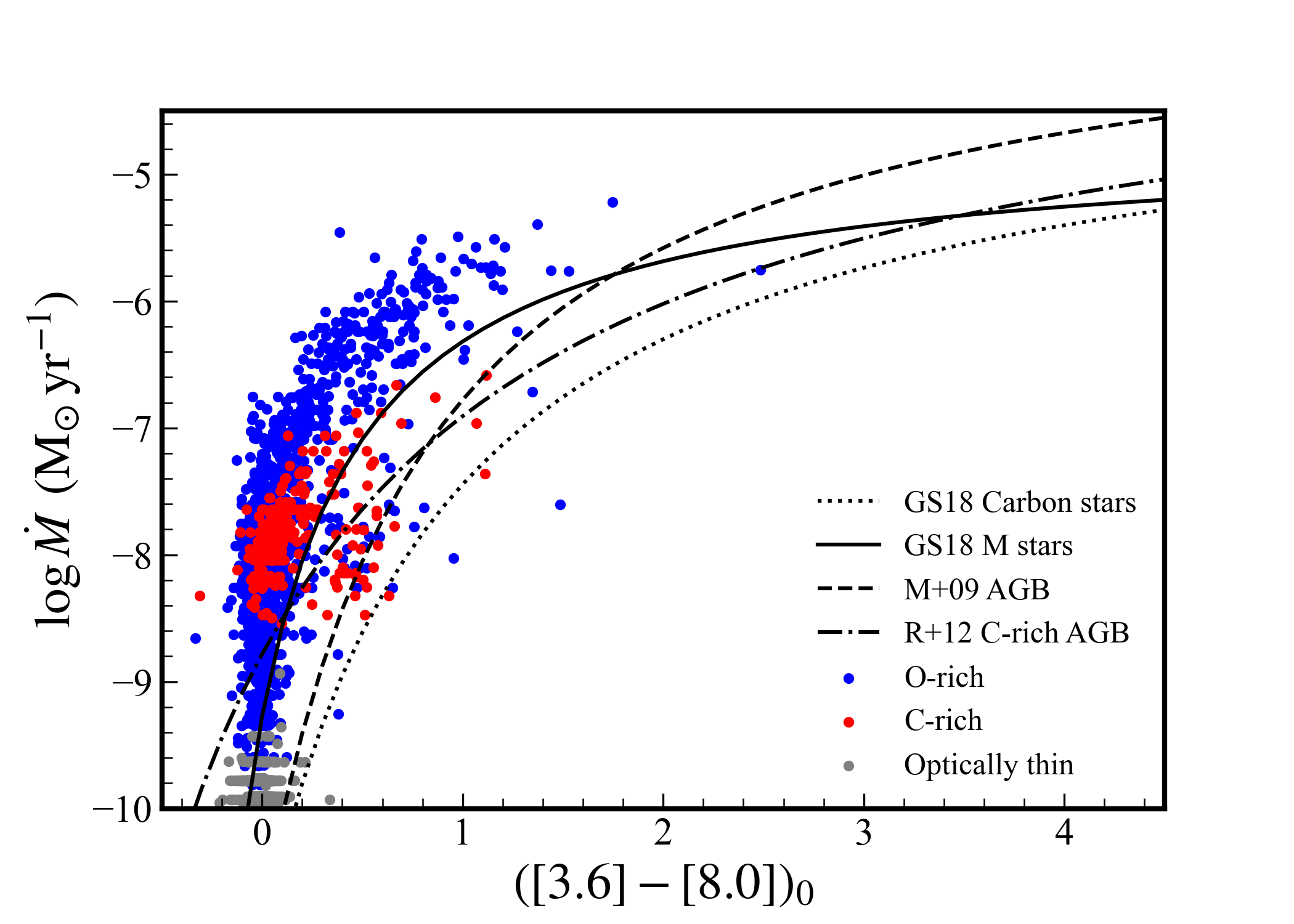}
    \caption{$\rm{[3.6]- [8.0]}$ color versus MLR diagram, we depicted the color - DPR relation obtained from previous work (black) too.}
    \label{fig:irac}
\end{figure}

We also investigated the relationship between DPR and stellar colors of $(J-\rm{[8.0]})_{0}$, $(K_{\rm{S}}-W3)_{0}$, $(\rm{[3.8]-[24]})_{0}$, as shown in Figure~\ref{fig:cmlr} (only RSGs with $sflag=0$ are included). Both the O-rich and C-rich RSGs exhibit positive correlations between the DPR and stellar infrared colors. Figure~\ref{fig:irac} shows our resulted $(\rm{[3.6]-[8.0]})_{0}$ color versus DPR diagram and the DPR-color relations from previous works. For comparison, we adopted a gas-to-dust ratio of 200 for both our results and the relations from the literature, in order to keep a consistent value of gas-to-dust ratio. The literature relations are mainly based on AGB and carbon stars. The RSGs show a roughly similar DPR-color relation. They are most similar to the M stars from GS18, but not so close to the carbon stars from R+12, \citet[M+09]{2009MNRAS.396..918M}, and GS18.

\section{Summary}
\label{sec:summary}
We establish a pre-computed library of evolved star templates using 2-DUST code \citep{2003ApJ...586.1338U}. Through SED fitting, the DPR and dust chemical properties of a complete and uncontaminated RSG sample of 4,714 RSGs in the LMC are determined. Our analysis reveal that the cumulative DPR of RSGs in the LMC is $1.14 \times 10^{-6}\, \rm{M_{\sun}\, yr^{-1}}$. We find that the DPR heavily depends on a few RSGs, as only 208 RSGs contribute 76.57\% of the DPR. Among the sample, 1,444 RSGs are classified as optically thin, meaning they have a minimal dust envelope. Despite accounting for 30.63\% of the total number, they only contributed 0.07\% of the DPR. Conversely, O-rich RSGs, accounting for 63.30\% of the total number, contributed 97.24\% of the DPR, making them the primary drivers of mass loss among RSGs. A small proportion of C-rich RSGs (6.07\%) contributed 2.69\% of total DPR. We verify our classifications with 86\% accuracy based on IRS spectrum.

We fit the luminosity-MLR relation and confirm the turning point in the relation at $\log{(L_{\rm{obs}}/L_{\sun})}\approx 4.4 $. Our result is consistent with that obtained by Y+23. Given the low mass loss rate derived in this work, whether RSGs can completely lose all their dust envelope is a question that requires further investigation.

Similar to AGB stars, we found a positive correlation between DPR of RSGs and infrared colors. Future work on AGB stars and other galaxies will determine the exact proportion of the contribution of RSGs and AGB to the dust of the whole galaxy and its own evolution.
                                                                            
\begin{acknowledgments}
 We are grateful to  Hai-Bo Yuan and Jia-Ming Liu for very helpful discussions. This work is supported by the National Key R\&D Program of China   No. 2019YFA0405500 and No. 2019YFA0405501, National Natural Science Foundation of China No. 12133002 and U2031209, 12203025, 12173034, 12373048 and 11833006, and Yunnan University grant No. C619300A034, Shandong Provincial Natural Science Foundation through project ZR2022QA064. We acknowledge the science research grants from the China Manned Space Project with No. CMS-CSST2021-A09, CMS-CSST-2021-A08 and CMS-CSST-2021-B03. The numerical computations were conducted on the Yunnan University Astronomy Supercomputer. This research made use of the cross-match service provided by CDS, TOPCAT \citep{2005ASPC..347...29T}.
\end{acknowledgments}

\bibliography{sample631}{}
\bibliographystyle{aasjournal}

\end{CJK*}
\end{document}